\documentclass[a4paper,11pt,oneside,fleqn]{article}

\usepackage[mathscr]{eucal}
\usepackage{amsmath,amsfonts,amssymb,amsthm,nccbbb,mathrsfs,verbatim}
\usepackage{graphicx}
\usepackage{caption}
\usepackage{subcaption}
\usepackage{color}
\usepackage[dvipsnames]{xcolor}
\usepackage{tikz,pgf}
\usetikzlibrary{patterns}
\usepackage{pdfpages}
\usepackage{soul}
\usepackage[ruled,vlined]{algorithm2e}
\usepackage{cancel}

\usepackage{tikz,pgf}
\usepackage{tikz-cd}
\usetikzlibrary{patterns}
\usepackage{standalone}
\usepackage{pgffor}

\usepackage{hyperref}
\usepackage[normalem]{ulem}
\hypersetup{colorlinks, linkcolor=blue, citecolor=blue, urlcolor=blue}
\allowdisplaybreaks
\numberwithin{equation}{section}
\usepackage[round]{natbib}
\definecolor{forestgreen(web)}{rgb}{0.13, 0.55, 0.13}

\newcommand{\delm}[1]{\ifmmode\text{\sout{\ensuremath{#1}}}\else\sout{#1}\fi}

\renewcommand*{\Omega}{\varOmega} 
\newcommand{\sub}{\scriptscriptstyle} 
\renewcommand*{\vec}{\boldsymbol} 
\newcommand{\df}{\mathrm{d}} 
\newcommand{\dt}{\df t} 
\newcommand{\sdbt}{\vec{\sigma} \df \vec{B}_t} 
\newcommand{\sdbth}{\vec{\sigma}_{\sub H} \df \vec{B}_t} 
\newcommand{\sdbtz}{\sigma_{\sub z} \df B_t} 
\newcommand{\dpt}{\df p_t^{\sigma}} 
\newcommand{\Df}{\mathrm{\mathbb{D}}} 
\newcommand{\Dfh}{\mathrm{\mathbb{D}}^{\sub H}}
\newcommand{\bdot}{\boldsymbol{\cdot}} 
\newcommand{\grad}{\boldsymbol{\nabla}} 
\newcommand{\gradh}{\boldsymbol{\nabla}_{\sub H}}
\newcommand{\tp}{^{\scriptscriptstyle T}} 
\newcommand{\defin}{\stackrel{\scriptscriptstyle\triangle}{=}} 
\newcommand{\gradp}{\boldsymbol{\nabla}^{\sub\perp}} 
\newcommand{\Exp}{\mathbb{E}} 
\newcommand{\tr}{\mathrm{tr}} 
\renewcommand*{\div}{\boldsymbol{\nabla\cdot}} 
\newcommand{\divh}{\boldsymbol{\nabla}_{\sub H}\boldsymbol{\cdot}}
\newcommand{\curl}{\boldsymbol{\nabla} \times} 
\newcommand{\adv}{\boldsymbol{\cdot\nabla}} 
\newcommand{\advh}{\bdot\boldsymbol{\nabla}_{\sub H}}
\newcommand{\laplac}{\nabla^{2}} 
\newcommand\Ro{\mbox{R}_{\mbox{o}}}  
\newcommand{\alf}{\frac{1}{2}} 
\newcommand{\real}{\mathbb{R}} 
\newcommand{\Id}{\vec{\mathrm{I}}} 
\newcommand{\KE}{\text{KE}}
\newcommand{\PE}{\text{PE}}
\newcommand{\E}{\text{E}}

\allowdisplaybreaks
\makeatletter
\long\def\@makecaption#1#2{%
  \vskip\abovecaptionskip\footnotesize
  \sbox\@tempboxa{#1. #2}%
  \ifdim \wd\@tempboxa >\hsize
    #1. #2\par
  \else
    \global \@minipagefalse
    \hb@xt@\hsize{\hfil\box\@tempboxa\hfil}%
  \fi
  \vskip\belowcaptionskip}
\makeatother

\newcommand{\todo}[1][\null]{\ensuremath{\clubsuit}}

\newcommand{\noprint}[1]{}
\newcommand{\checked}[1][\null]{\ensuremath{\boldsymbol{\surd}}}

\newtheorem*{problem*}{Problem}
{\theoremstyle{definition}

\newtheorem*{remark*}{Remark}
}

\begin{document}

\par\noindent {\LARGE\bf
Rotating shallow water flow under location uncertainty with a structure-preserving discretization
\par}

{\vspace{4mm}\par\noindent {\bf 
R\"{u}diger Brecht$^{\dag}$, Long Li$^{\S}$, Werner Bauer$^{\ddag}$, Etienne M\'{e}min$^{\S}$
} \par\vspace{2mm}\par}

{\vspace{2mm}\par\noindent {\it
		$^{\dag}$~Department of Mathematics and Statistics, Memorial University of Newfoundland,\\ St.\ John's (NL) A1C 5S7, Canada
}}
{\vspace{2mm}\par\noindent {\it
		$^{\S}$~Inria/IRMAR, Campus universitaire de Beaulieu, Rennes, France

}}
{\vspace{2mm}\par\noindent {\it
		$^{\ddag}$~Imperial College London, Department of Mathematics, 180 Queen’s Gate, London SW7 2AZ, United Kingdom.
}}

\vspace{4mm}\par\noindent\hspace*{10mm}\parbox{140mm}{\small

}\par\vspace{4mm}


\section*{Abstract}
We introduce a physically relevant stochastic representation of the rotating shallow water equations. The derivation relies mainly on a stochastic transport principle and on a decomposition of the fluid flow into a large-scale component and a noise term that models the unresolved  flow components. As for the classical (deterministic) system, this scheme, referred to as modelling under location uncertainty (LU), conserves the global energy of any realization and provides the possibility to generate an ensemble of physically relevant random simulations with a good trade-off between the model error representation and the ensemble's spread.  To maintain numerically the energy conservation feature, we combine an energy (in space) preserving discretization of the underlying deterministic model with approximations of the stochastic terms that are based on standard finite volume/difference operators. The LU derivation, built from the very same conservation principles as the usual geophysical models, together with the numerical scheme proposed  can be directly used in existing dynamical cores of global numerical weather prediction models. The capabilities of the proposed framework is demonstrated for an inviscid test case on the f-plane and for a barotropically unstable jet on the sphere.

\section*{Plain summary}

The motion of geophysical fluids on the globe needs to be modelled to get insights of tomorrow's weather. These forecasts must be precise enough while remaining computationally affordable. Ideally they should enable to estimate likely scenarios through an ensemble of physically relevant realizations, built from an accurate handling of the model errors that are inescapably introduced due to physical or numerical approximations. To address these issues, we advocate the use of a stochastic framework to represent the action of the many unresolved fast/small-scale processes on the resolved flow component. 
The derivation  of the stochastic system, based on the usual conservation laws, is presented in detail and simulated with an adapted structure preserving numerical model to maintain numerically the nice properties of the stochastic setting inherited from a transport principle, namely: mass and energy conservation. 
The versatile nature of the stochastic derivation as well as of the proposed numerical scheme makes this framework suitable for existing dynamical cores of global numerical weather prediction models. Numerical results illustrate the energy conservation of the numerical model and the accuracy of large-scale stochastic simulations when compared to corresponding deterministic ones. The ability of the random dynamical system to represent model errors is also shown.

\section{Introduction}

Numerical simulations of the Earth's atmosphere and ocean play an important role in developing our understanding of weather forecasting.
A major focus lies in determining the large-scale flow correctly, which is strongly related to the parameterizations of sub-grid processes \cite{frederiksen2013}. The non-linear and non-local nature of the dynamics of geophysical fluid flows make the large-scale flow structures interact with the smaller components. Solving the Kolmogorov scales \cite{Pope2000} of geophysical flows is today, and likely for a foreseeable future, completely out of reach. 
This is due, in the first place, to the formidable computational expense that would be necessary, but also to the complexity of the many fine-scale physical or bio-chemical processes involved. Truncating the fine scales and simply ignoring their actions is highly detrimental to a reliable simulation of the large-scale components of the flow.  Yet, an accurate modelling of the fine-scale processes' effects is an excruciatingly difficult  task and the idea of a stochastic modelling has strongly attracted the geophysical community since the seminal works of \cite{Hasselmann-76} and \cite{Leith75}. For several years, this interest has been strongly strengthened with the emergence of ensemble methods for probabilistic forecasting and data assimilation issues \cite{Berner-17, Franzke2015, Gottwald2017, Majda2008, Palmer08, Slingo11}.  

The schemes proposed so far rely on very different methodological concepts. Multiplicative random forcing and randomization of parameters based on early turbulence studies on energy backscattering \cite{Leith90,Mason92} have been proposed \cite{Buizza99, PortaMana-14,Shutts05}. 
The ad hoc nature of these schemes makes a systematic stochastic derivation of any flow dynamical model or configuration difficult. In addition, the absence of an explicit energy balance of the noise term leads to an uncontrolled increase of variance that is potentially problematic. 
They consequently require a proper tuning of the large-scale sub-grid model and of the noise amplitude to stabilize the system. 
The subgrid model is, however, not related to the noise term and the amplitude of the perturbations to apply is also difficult to specify on physical grounds. More importantly, even for low noise, an arbitrary random perturbation defined outside of the physical principles on which the system has been built upon may lead to strongly erroneous probability density functions of the system's dynamics  \cite{Chapron2018}. Other schemes based on an averaging and homogenization theory have been proposed  \cite{Franzke2005, Franzke2006}  in the wake of \cite{Majda99} and extended through the Mori-Zwanzig formalism (see the review \cite{Gottwald2017} and references therein). Those techniques are well suited for the design of stochastic reduced order systems.

In this study, we propose to stick to a specific stochastic model, called modelling under \emph{Location Uncertainty} (LU) derived by \cite{Memin2014}, which emerges from a decomposition of the Lagrangian velocity into a smooth-in-time drift and a highly oscillating random term. Such a slow/fast or smooth/oscillating decomposition is reminiscent to the Lagrangian decomposition introduced in the seminal work of \cite{Andrews78}, which is currently used for surface or internal waves studies \cite{kafiabad_vanneste_young_2021,Salmon2013, Young97, Xie2015}. A similar random decomposition is also at the center of the variational stochastic framework of \cite{Holm-15}. Like our setting this latter approach applies in a broader context and not only to wave solutions. Both frameworks rely on a stochastic transport principle, with  \cite{Holm-15} dedicated to Hamiltonian dynamical systems and defined from a circulation preserving constrained variational formulation, while \cite{Memin2014} is general and built upon classical physical conservation laws.

This stochastic transport principle has been used as a fundamental tool to derive stochastic representations of large-scale geophysical dynamics \cite{Bauer2020jpo, Bauer2020ocemod, Chapron2018, Resseguier2017gafd1,Resseguier2017gafd2, Resseguier2017gafd3} or to define large eddy simulation models of turbulent flows \cite{Chandramouli-JCP-20, Kadri2017}. The LU framework relies on a stochastic representation of the Reynolds transport theorem \cite{Kadri2017,Memin2014} which introduces naturally meaningful terms for turbulence studies. 

It gathers a multiplicative random advection which is responsible for an energy backscattering, a subgrid diffusion operator describing the mixing of the large-scale flow component by the small-scale random component, and an effective advection which is attached to the small scales spatial inhomogeneity. This latter term has been shown to be reminiscent of a generalized Stokes drift component, hence designated as It\^{o}-Stokes drift \cite{Bauer2020jpo}. Backscattering and diffusion are energetically in balance which leads hence to global energy conservation. 

Recently, the LU formulation was shown to perform very well for oceanic quasi-geostrophic flow models \cite{Resseguier2017gafd2, Resseguier2017gafd3, Bauer2020jpo, Bauer2020ocemod}. It was found to be more accurate in predicting the extreme events, in diagnosing the frontogenesis and filamentogenesis, in structuring the large-scale flow and in reproducing long-terms statistics. Besides, for a LU version of the Lorentz-63 model, derived from a Rayleigh-B\'{e}nard convection in the very same way as the original model  \cite{Berge-Pomeau-Vidal, Lorenz63}, it has been demonstrated that the LU setting was more effective in exploring the range of the strange attractor compared to classical models as well as to stochastic models built with ad hoc multiplicative forcings \cite{Chapron2018}. 

In this work, the performance of the LU representation is assessed for the numerical simulation of the rotating shallow water (RSW) system, which can be considered as the first step towards developing global random numerical weather prediction and climate models. In particular, this is the first time that the LU formulation is implemented for the dynamics evolving on the sphere. 
The global energy conservation of the RSW-LU system for any realization, which is analytically demonstrated here, is a strong asset of the approach and this invariant feature should  be numerically conserved as closely as possible.  
Global energy conservation is especially  important for long-term climatic simulations.	
However, classical purely damping parameterizations do not take into account energy and momentum fluxes from the unresolved to the resolved scales. In climatic models, this is believed to be a source of important biases  \cite{Gugole2019}. 

Hence, we propose to combine the discrete variational integrator for RSW fluids as introduced in \cite{Bauer2019} and \cite{Brecht2019}
with the numerical LU setting in order to maintain this conservation property as well as all the transport invariants. 
The benefit of the proposed method that relies on a modular combination of a variational integrator with a (potentially different) discretization of the LU formulation is that it should be directly applicable to 
existing dynamical cores of numerical weather prediction models.

The derivation of the variational integrator is based on the variational discretization framework introduced by \cite{pavl11a} for incompressible fluids, expanded by \cite{gawlik2011geometric} 
to incompressible fluids with advected quantities. In various papers, this framework has been further 
extended, for instance \cite{desbrun2014variational} incorporated rotating and stratified fluids of atmospheric and oceanic dynamics and \cite{Bauer2019a} introduced soundproof approximations of the Euler equations. 
Variational integrators are designed by first discretizing the given Lagrangian, and then by deriving a discrete system of associated Euler-Lagrange equations from the discretized Lagrangian (see \cite{marsden2001discrete}).

The advantage of this approach is that the resulting discrete system inherits several important properties of the underlying 
continuous system, notably a discrete version of Noether's theorem that guarantees the preservation of conserved quantities associated to the symmetries of the discrete Lagrangian (see \cite{hairer2006geometric}). 
Variational integrators also exhibit superior long-term stability properties, cf. e.g. \cite{leim04Ay}.
Therefore, they typically
outperform traditional integrators if one is interested in long-time integration or the statistical properties of a given dynamical system. Our choice for an energy preserving rather than an enstrophy 
conserving scheme is based on the following considerations. 
As shown in \cite{Bauer2020ocemod} for stochastic barotropic quasi-geostrophic models, using an energy conserving scheme for long-term predictions yields better results than using an enstrophy conserving one.
Besides, because of the direct cascade of enstrophy to high wave numbers, often stabilization through enstrophy dissipation is introduced, even in initially enstrophy conserving schemes, cf.  \cite{bonaventura2005analysis,mcrae2014energy,ringler2002potential}.


Apart from taking into account the unresolved processes, it is paramount in probabilistic ensemble forecasting to model the  uncertainties along time \cite{Resseguier2020arcme}. In particular, operational ensemble data assimilation  methods rely classically on random perturbations of the initial conditions (PIC) together with an artificially carefully inflated variance  \cite{Anderson1999} to increase the otherwise deficient  ensemble forecasts' spread \cite{Gottwald2013, Franzke2015}. Such inflation has the side effect of augmenting also the representation error of the ensemble members. 
In the present work, we compare the reliability of the ensemble spread of such a PIC model with our RSW-LU system, under the same noise amplitude, and show that the LU strategy yields a good trade-off between model error representation and ensemble spread.

The remainder of this paper is structured as follows. Section \ref{sec:RSW-LU} describes the basic principles of the derivation of the rotating shallow water system in the  LU formulation. 
Section \ref{section-discLU} explains 
the numerical discretization of the stochastic dynamical system. 
Section \ref{sec:results} discusses the numerical results for an inviscid test case with homogeneous noise and a viscous test case with heterogeneous noise. In Section~\ref{sec:conclusions} we draw some conclusions and provide an outlook for future work.
In the Appendices we demonstrate the energy conservation of the RSW--LU system, review some parameterizations of the noise and describe the discretization of the stochastic terms.


\section{Rotating shallow water equations under location uncertainty}\label{sec:RSW-LU}

In this section, we first review the LU representation introduced by \cite{Memin2014}, then we derive the rotating shallow water equations under LU, denoted as RSW--LU, following the classical strategy  \cite{Vallis2017}. In particular, we demonstrate one important characteristic of the RSW--LU, namely that it preserves the total energy of the large-scale flow.

\subsection{Location uncertainty principles}

The LU formulation is based on  a temporal-scale-separation assumption of the following stochastic flow:
	\begin{equation}\label{eq:dX}
	\df \vec{X}_t = \vec{w}(\vec{X}_t, t)\, \dt + \vec{\sigma} (\vec{X}_t, t)\, \df \vec{B}_t,
	\end{equation}
where $\vec{X}$ is the Lagrangian displacement defined within the bounded domain $\Omega \subset \real^d\ (d = 2\ \text{or}\ 3)$, $\vec{w}$ is the large-scale velocity that is both spatially and temporally correlated, and $\sdbt$ is a highly oscillating  unresolved component (also called noise) term that is only correlated in space. The spatial structure of such noise is specified through a deterministic integral operator $\vec{\sigma}: (L^2 (\Omega))^d \rightarrow (L^2 (\Omega))^d$, acting on square integrable vector-valued functions $\vec{f}\in (L^2 (\Omega))^d$, with a bounded kernel  $\breve{\vec{\sigma}}$ such that 
	\begin{equation}\label{eq:corr}
	\vec{\sigma} [\vec{f}] (\vec{x}, t)  = \int_{\Omega} \breve{\vec{\sigma}} (\vec{x}, \vec{y}, t) \vec{f} (\vec{y})\, \df \vec{y},\ \quad \forall \vec{f} \in (L^2 (\Omega))^d.
	\end{equation}
The randomness of such a noise is driven by a functional Brownian motion $\vec{B}_t$ \cite{DaPrato2014}.
The fact that the kernel is bounded, 
implies that the resulting random flow $\sdbt$ is a centered (of null ensemble mean) Gaussian process with the well-defined \emph{covariance tensor}:

	\begin{align}\label{eq:cov}
	\vec{Q} (\vec{x}, \vec{y}, t, s) &= \Exp \Big[ \big( \vec{\sigma} (\vec{x}, t)\, \df \vec{B}_t \big) \big( \vec{\sigma} (\vec{y}, s)\, \df \vec{B}_s \big)\tp \Big] \nonumber \\
	&= \delta(t-s)\, \dt \int_{\Omega}{\breve{\vec{\sigma}} (\vec{x}, \vec{z}, t) \breve{\vec{\sigma}}\tp (\vec{y}, \vec{z}, s)}\, \df \vec{z},
	\end{align}

where $\Exp$ stands for the expectation, $\delta$ is the Kronecker symbol and $\bullet\tp$ denotes matrix or vector transpose. The strength of the noise is measured by its \emph{variance}, denoted here as $\vec{a}$, and which is  given by the diagonal components of the covariance per unit of time:

	\begin{equation}\label{eq:var0}
	\vec{a} (\vec{x}, t) \dt = \vec{Q} (\vec{x}, \vec{x}, t, t).
	\end{equation}

We remark that this  variance tensor has the same unit as a diffusion tensor ($\text{m}^2 \cdot \text{s}^{-1}$) and that the density of the turbulent kinetic energy (TKE) can be specified through it by $\alf \tr (\vec{a}) / \dt$.

The previous representation \eqref{eq:corr} is a general way to define the noise, but other formulations can be  conveniently used in practice. In particular, the covariance operator per unit of time, $\vec{Q}/\dt$, admits an orthogonal eigenfunction basis $\{  \vec{\Phi}_n (\bullet, t) \}_{\sub n \in \mathbb{N}}$ weighted by the eigenvalues $\Lambda_n \geq 0$ such that $\sum_{\sub n \in \mathbb{N}} \Lambda_n < \infty$. Therefore, one may equivalently define the noise and its variance, based on the following spectral decomposition:

	\begin{equation}\label{seq:KL}
	\vec{\sigma} (\vec{x}, t)\, \df \vec{B}_t = \sum_{n \in \mathbb{N}} \vec{\Phi}_n (\vec{x}, t)\, {\df \beta_t^n},\ \quad \vec{a} (\vec{x}, t) = \sum_{n \in \mathbb{N}} \vec{\Phi}_n (\vec{x}, t) \vec{\Phi}_n\tp (\vec{x}, t),
	\end{equation}

where $\beta^n$ denotes $n$ independent and identically distributed (i.i.d.) one-dimensional standard Brownian motions. The specification of those basis functions from data driven empirical covariance matrices enables one to construct specific noises, informed either by numerical or observational data. This strategy will allow us to devise various forms of the noise in the following.

\noindent\paragraph*{Remark 1} Decomposition \ref{eq:dX} is a temporal decomposition and not a spatial decomposition as classically formulated through spatial filters and/or decimation operators in large-eddies simulation (LES) techniques. However, in the case of  turbulent flows, time and spatial scales are related. As a matter of fact, in the inertial range, the turn-over time ratio for two different scales $L$ and $\ell$ reads  $\tau_L/\tau_{\ell} \propto (L/\ell)^{2/3}$ and provides a direct relation between time-scale coarsening and spatial-scale dilation.  Unless specifically needed, in the following, we will thus refer to large/small or unresolved scales without differentiating between time or space scales. Note also that temporal filtering has already been used for the definition of oceanic models \cite{Hecht08} or large-eddies simulation approaches \cite{Meneveau00}.

\paragraph*{Remark 2} Decomposition \ref{eq:dX} is written in terms of an It\^{o} stochastic integral. This decomposition could have been written in the form of a Stratonovich integral as well. The calculus associated to this latter integral has the advantage of following the classical  chain rule. However, the Stratonovich noise no longer has zero expectation.  This leads thus to a problematic decomposition with velocity fluctuations of non null ensemble mean. For smooth enough integrands, it is possible to safely move from one form to the other.  For interested readers, more insights on the difference of the two settings and their implications in stochastic oceanic modelling are provided in \cite{Bauer2020jpo}.

\paragraph*{Remark 3} The approach could be extended to express flows on arbitrary Riemannian manifolds. In that case it is  easier to work directly with the Stratonovich formulation since it is invariant under the change of coordinates.
As we consider here only flows that assume the shallow approximation, the considered representation of the equations in $\real^2$ and $\real^3$ is a very accurate approximation.


\bigskip

The core of the LU model representation is based on a stochastic Reynolds transport theorem (SRTT), introduced by \cite{Memin2014}, which describes the rate of change of a random scalar $q$ transported by the stochastic flow \eqref{eq:dX} within a flow volume $\mathcal{V}$. In particular, for incompressible unresolved flows, $\div \vec{\sigma} = 0$, the SRTT can be written as

	\begin{subequations}
		\begin{align}
		&\df_t\, \Big( \int_{\mathcal{V}(t)} q (\vec{x}, t)\, \df \vec{x} \Big) = \int_{\mathcal{V}(t)} \big( \Df_t q + q \div (\vec{w} - \vec{w}_s) \big)\, \df \vec{x}, \label{eq:SRTT} \\
		&\Df_t q = \df_t q + (\vec{w} - \vec{w}_s) \adv q\, \dt + \sdbt \adv q - \alf \div (\vec{a} \grad q)\, \dt, \label{eq:STO} 
		\end{align}
	\end{subequations}

where $\df_t q (\vec{x}, t) = q (\vec{x}, t + \dt)  - q (\vec{x}, t)$ stands for the forward time-increment of $q$ at a fixed point $\vec{x}$, $\Df_t$ is introduced as the stochastic transport operator in \cite{Resseguier2017gafd1} and $\vec{w}_s = \alf \div \vec{a}$ is referred to as the It\^{o}-Stokes drift (ISD) in \cite{Bauer2020jpo}. The transport operator plays the role of the material derivative in the stochastic setting. The ISD is defined by the variance tensor divergence and embodies the effect of statistical inhomogeneity of the unresolved flow on the large-scale component. As shown in  \cite{Bauer2020jpo}, it can be considered as a generalization of the Stokes drift associated to waves propagation  with the emergence of a similar vortex force and Coriolis correction. In the definition of the stochastic transport operator in \eqref{eq:STO}, 
the last two terms describe, respectively, an energy backscattering from the unresolved scales to the large scales and an inhomogeneous diffusion of the large scales driven by the variance of the unresolved flow components. The diffusion term generalizes the Boussinesq eddy viscosity assumption (here with a matrix eddy viscosity). This term is, nevertheless, directly related to the noise form and not anymore defined by loose analogy with the molecular dissipation mechanism. The backscattering term corresponds to an energy  source that is exactly compensated by the diffusion term \cite{Resseguier2017gafd1}. 

In particular, for an isochoric flow with $\div (\vec{w} - \vec{w}_s) = 0$, one may immediately deduce from \eqref{eq:SRTT} the following transport equation of an extensive scalar:

	\begin{equation}
	\Df_t q = 0,
	\end{equation}

where the energy of such random scalar $q$ is globally conserved, as shown in \cite{Resseguier2017gafd1}:

	\begin{equation}\label{eq:energy-balance}
	\df_t\, \Big( \int_{\sub\Omega} \alf q^2\, \df\vec{x} \Big) = \Big( \underbrace{\alf\int_{\sub\Omega} q \div (\vec{a} \grad q)\, \df\vec{x}}_{\text{Energy loss by diffusion}} + \underbrace{\alf\int_{\sub\Omega} (\grad q)\tp \vec{a} \grad q\, \df\vec{x}}_{\text{Energy intake by noise}} \Big)\, \dt = 0.
	\end{equation}

Indeed, this can be interpreted  as a process where the energy brought by the noise is exactly counterbalanced by that dissipated from the diffusion term.

\subsection{Derivation of RSW--LU}\label{appendix-RSWLU}

This section describes in detail the derivation of the RSW--LU system. This model enriches the formulation described in \cite{Memin2014}. Here it is fully stochastic and includes rotation to suit simulations of geophysical flows on a rotating frame.

The above SRTT \eqref{eq:SRTT} and Newton's second principle allow us to derive the following (three-dimensional) stochastic equations of motions in a rotating frame \cite{Bauer2020jpo}:

	\begin{subequations}
		\begin{align}
		&{\text{\em Horizontal momentum equation}}:\nonumber\\
		&\Df_t \vec{u} + \vec{f} \times \big( \vec{u}\, \dt + \sdbth \big) = - \frac{1}{\rho} \gradh \big( p\, \dt + \dpt \big) + \nu \laplac \big( \vec{u}\, \dt + \sdbth\big), \label{eq:hmoment1}\\
		&{\text{\em Vertical momentum equation}}:\nonumber\\
		&\Df_t w = - \frac{1}{\rho} \partial_z \big( p\, \dt + \dpt \big) - g\, \dt + \nu \laplac \big( w\, \dt + \sdbtz \big), \label{eq:vmoment1}\\
		&{\text{\em  Mass equation}}:\nonumber\\
		&\Df_t \rho = 0, \label{eq:mass1} \\
		&{\text{\em Continuity equations}}:\nonumber\\
		&\divh \big( \vec{u} - \vec{u}_s \big) + \partial_z (w - w_s) = 0,\ \quad \divh \sdbth + \partial_z \sdbtz = 0, \label{eq:continu1} 
		\end{align}
	\end{subequations}

where $\vec{u}= (u,v)\tp$ (resp. $\sdbth$) and $w$ (resp. $\sdbtz$) are the horizontal and vertical components of the three-dimensional large-scale flow $\vec{w}$ (resp. the unresolved random  flow $\sdbt$); $\vec{f} = (2 \tilde{\Omega} \sin \Theta) \vec{k}$ is the Coriolis parameter varying in latitude $\Theta$, with the Earth's angular rotation rate $\tilde{\Omega}$ and the vertical unit vector $\vec{k} = [0, 0, 1]\tp$; $\rho$ is the fluid density; $\gradh = [\partial_x, \partial_y]\tp$ denotes the horizontal gradient; $p$ and $\dot{p}_t^{\sigma} = \dpt / \dt$ (informal definition) are the time-smooth and time-uncorrelated components of the pressure field, respectively; $g$ is the Earth's gravity value and $\nu$ is the kinematic viscosity. In the following, the molecular friction term is assumed to be negligible and dropped from the equations. Note that in our setting the continuity equations \eqref{eq:continu1} ensure volume conservation \cite{Resseguier2017gafd1} and mass conservation \eqref{eq:mass1}.

In order to model the large-scale circulations in the atmosphere and ocean, the hydrostatic balance approximation is widely adopted \cite{Vallis2017}. We now specify the scaling for this balance in the LU framework. We first adimensionalize the basic variables as  

	\begin{equation}
	(x, y) = \mathcal{L}\, (x', y'),\ \quad \vec{u} = \mathcal{U}\, \vec{u}',\ \quad t = \mathcal{T}\, t',\ \mathcal{T} = \mathcal{L} / \mathcal{U},\ \quad z = \alpha \mathcal{L} z',\ \alpha = \mathcal{H} / \mathcal{L},  
	\end{equation}

where the capital letters are used for the characteristic scales of variables and $\bullet'$ denotes adimensional variables. To scale properly the vertical velocity, we propose to adopt a sufficient incompressible condition \cite{Resseguier2017gafd1,Resseguier2017gafd2} for the resolved component in Equation \eqref{eq:continu1}, that is

	\begin{equation}\label{eq:continu2}
	\divh \vec{u} + \partial_z w = 0,\ \quad \divh \vec{u}_s + \partial_z w_s = 0.
	\end{equation}

Note that the latter divergence-free condition on the ISD is usually considered for the classical Stokes drift \cite{Mc-Williams-Restrepo-Lane-04} although being controversial \cite{Mellor2016}. The three-dimensional bolus velocity introduced in the eddy-induced-advection parametrization \cite{Gent1990,Gent1995,Griffies1998} is also assumed to be incompressible in order to preserve the tracer's moments.
In our case, the justification of this constraint is further strengthen by global energy conservation and a desirable bridge between the classical (global energy conserving) rotating shallow water system and its stochastic representation. Under the condition \eqref{eq:continu2}, a classical scaling of the vertical (resolved) velocity holds:

	\begin{equation}
	w = \alpha\, \mathcal{U}\, w'.
	\end{equation}

Apart from these classical scaling numbers, the horizontal component $\vec{a}_{\sub H}$ of the variance/diffusion tensor $\vec{a}$, which characterizes the strength of the unresolved component, is scaled as

	\begin{equation}\label{eq:decom-var1}
	\vec{a}_{\sub H}  = \epsilon\, \mathcal{UL}\, \vec{a}_{\sub H}',\ \quad \vec{a} = 
	\begin{pmatrix}
	\vec{a}_{\sub H}   & \vec{a}_{\sub Hz} \\
	\vec{a}_{\sub Hz} & a_{\sub z}
	\end{pmatrix},\ \quad
	\epsilon = \frac{\mathcal{T}_{\sigma}}{\mathcal{T}} \frac{\text{EKE}}{\text{MKE}},
	\end{equation}

where the specific factor $\epsilon$ \cite{Resseguier2017gafd2} is defined as the ratio between the eddy kinetic energy (EKE) and the mean kinetic energy (MKE), multiplied by the ratio between the unresolved scale correlation time $\mathcal{T}_{\sigma}$ and the large-scale advection time. From the definitions \eqref{eq:cov} and \eqref{eq:var0}, the scaling of the horizontal small-scale flow reduces to

	\begin{equation}
	\sdbth = \sqrt{\epsilon}\, \mathcal{L}\, (\sdbth)'.
	\end{equation}

In addition, we consider the following scaling between  the vertical and horizontal components of the unresolved flow:

	\begin{equation}
	\frac{\sdbtz}{\|\sdbth\|} \sim \alpha\, \delta,\ \quad \text{i.e.}\ \sdbtz = \sqrt{\epsilon}\, \delta\, \mathcal{H}\, (\sdbtz)',
	\end{equation}

where $\delta$ is a small factor \cite{Resseguier2017gafd2}. Again, from the definitions \eqref{eq:cov} and \eqref{eq:var0}, the other components of the variance/diffusion tensor scale then as:

	\begin{equation}
	\vec{a}_{\sub Hz} = \epsilon\, \delta\, \mathcal{UH}\, \vec{a}_{\sub Hz}',\ \quad a_{\sub z} =  \epsilon\, \delta^2\, \alpha\, \mathcal{UH}\, a_{\sub z}',\ \quad \quad \text{i.e.}\ \frac{a_{\sub z}}{\|\vec{a}_{\sub H}\|} \sim \alpha^2 \delta^2.
	\end{equation}

This  relation provides a ratio between the vertical and  horizontal eddy diffusivities. It is in practice quite small at large scale \cite{Levy2010,Levy2012}.

Now, with $\vec{f} = 0$ and a constant density $\rho_0$, the horizontal momentum equation \eqref{eq:hmoment1} implies the following scalings of the rescaled pressures:

	\begin{equation}
	\tilde{p} = p / \rho_0 = \mathcal{U}^2\, \tilde{p}',\ \quad \df \tilde{p}_t^{\sigma} = \dpt / \rho_0 = \sqrt{\epsilon}\, \mathcal{U L}\, (\df \tilde{p}_t^{\sigma})'.
	\end{equation}

Finally, substituting all the above scalings into Equation \eqref{eq:vmoment1}, the adimensional vertical momentum is given by

	\begin{align}\label{eq:vmoment2}
	\alpha^2\, \bigg[ \df_t w' &+ (\vec{u}' \advh' w' + w' \partial_z' w')\, \dt' + \sqrt{\epsilon} \big( (\sdbth)' \advh' w' + \delta\, (\sdbtz)' \partial_z' w' \big)\nonumber \\
	&- \frac{\epsilon}{2} \Big( (\gradh' \bdot \vec{a}_{\sub H}' + \delta\, \partial_z' \vec{a}_{\sub Hz}') \advh' w' + \delta\, (\gradh' \bdot \vec{a}_{\sub Hz}' + \delta\, \partial_z' a_{\sub z}') \partial_z' w'  \nonumber \\
	&{}\ {}\ {}\ {}\ {}\ + \gradh' \bdot (\vec{a}_{\sub H}' \gradh' w' + \delta\, \vec{a}_{\sub Hz}' \partial_z' w') + \delta\, \partial_z' (\vec{a}_{\sub Hz}' \gradh' w' + \delta\, a_{\sub z}' \partial_z' w') \Big)\, \dt' \bigg] \nonumber \\
	&= - \partial_z' \big( \tilde{p}'\, \dt' + \sqrt{\epsilon}\, (\df \tilde{p}_t^{\sigma})' \big) - \frac{1}{\text{Fr}^2}\, \dt',
	\end{align}

where $\text{Fr} = \mathcal{U} / \sqrt{g \mathcal{H}}$ is the \emph{Froude number}. Let us now make the following assumptions:

	\begin{equation}
	\alpha^2 \ll 1,\ \quad \text{Fr}^2 = \mathcal{O} (1),\ \quad \epsilon = \mathcal{O} (1),\ \quad \delta \ll 1.
	\end{equation}

The acceleration term on the left-hand side (LHS) of Equation \eqref{eq:vmoment1} has now a lower order of magnitude than the RHS terms. Restoring the dimensions, the hydrostatic balance under moderate horizontal uncertainty and weak vertical uncertainty hence boils down to
\begin{subequations}
	
		\begin{equation}\label{eq:hydro1}
		\partial_z \big( p\, \dt + \dpt \big) = - \rho g\, \dt,\ \quad \text{i.e.}\ \partial_z p = - \rho g,\ \partial_z \dpt = 0.
		\end{equation}
	
	We remark that the unique decomposition principle of a semimartingale process \cite{Kunita1997} is used here to separate the bounded variation component (in terms of $\dt$) and the martingale part (in terms of $\df \vec{B}_t$ or $\dpt$).
	\begin{figure}[h]
		\centering
		\includegraphics[width=0.65\textwidth]{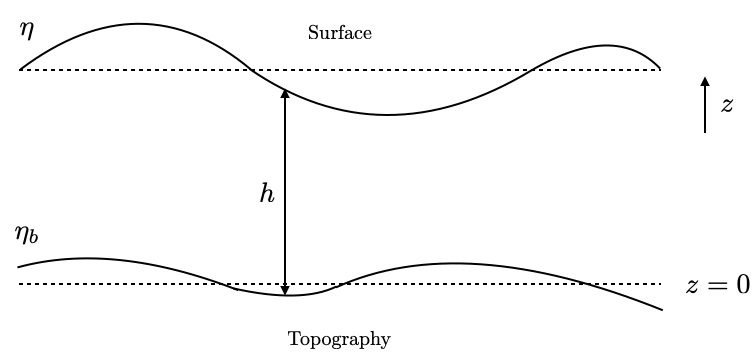}
		\caption{Illustration of a single-layered shallow water system (inspired by \cite{Vallis2017}). $h$ is the thickness of a water column, $\eta$ is the height of the free surface and $\eta_b$ is the height of the bottom topography. As a result, we have $h = \eta - \eta_b$.}
		\label{fig:RSW}
	\end{figure}
	Integrating vertically these hydrostatic balances \eqref{eq:hydro1} from $0$ to $z$ (see Figure \ref{fig:RSW}), we have
	
		\begin{align}\label{eq:p1} 
		p (x,y,z,t) = p_0(x,y,t) - \rho_0 g z,\ \quad
		\dpt (x,y,z,t) = \dpt(x,y,0,t),
		\end{align}
	
	where $p_0$ denotes the pressure at the bottom of the basin $( z = 0 )$. Following \cite{Vallis2017}, we assume that the weight of the overlying fluid is negligible, {\em i.e.} $p(x,y,\eta,t) \approx 0$ with $\eta$ the height of the free 
	surface, leading to $p_0 = \rho_0 g \eta$. This allows us to rewrite Equation~\eqref{eq:p1}
	such that for any $z \in [0, \eta]$ we have 
	
		\begin{equation}\label{eq:pl1}
		p (x,y,z,t) = \rho_0 g \big( \eta (x,y,t) - z \big).
		\end{equation}
	
	Subsequently, the pressure gradient force in the horizontal momentum equation \eqref{eq:hmoment1} reads
	
		\begin{equation}
		- \frac{1}{\rho_0} \gradh \big( p\, \dt + \dpt \big) = - g \gradh \eta - \frac{1}{\rho_0} \gradh \dpt,
		\end{equation}
	
\end{subequations}
which does not depend on $z$ according to Equations \eqref{eq:p1} and \eqref{eq:pl1}. Therefore, the acceleration terms on the LHS of Equation \eqref{eq:hmoment1} cannot depend on $z$, and the shallow water momentum equation under weak vertical uncertainty ($\delta \ll 1$) can be written finally as
\begin{subequations}
	
		\begin{align}
		&\Dfh_t \vec{u} + \vec{f} \times \big( \vec{u}\, \dt + \sdbth \big) = - g\gradh\eta\, \dt - \frac{1}{\rho_0} \gradh \dpt, \label{eq:RSW-moment} \\
		&\Dfh_t u = \df_t u + \big( (\vec{u} - \vec{u}_s)\, \dt + \sdbth \big) \advh u - \alf \divh \big( \vec{a}_{\sub H} \gradh u \big)\, \dt, \label{eq:Dt-ito-u} 
		\end{align}
	
\end{subequations}
where  $\vec{u}_s =  \alf \divh \vec{a}_{\sub H}$ is the two-dimensional ISD and $\Dfh_t$ denotes the horizontal stochastic transport operator whose expression is recalled in \eqref{eq:Dt-ito-u} for the $u$ component. The relation between the unresolved flow component and the random pressure can be further specified by considering a scaling of the martingale part of the momentum equation:

	\begin{equation}
	\sqrt{\epsilon}\, \df_t \tilde{u}' + \sqrt{\epsilon}\, (\sdbth)' \advh' u' +  \frac{\sqrt{\epsilon}}{\Ro}\, \vec{f}' \times (\sdbth)'= \sqrt{\epsilon}\,\gradh' (\dpt)',
	\end{equation}

where $\Ro = \mathcal{U} / (f_0 \mathcal{L})$ denotes the Rossby number with $\vec{f} = f_0 \vec{f}'$, and $\tilde{u} = u - \Exp(u)$ stands for the martingale part of the horizontal velocity. We note that the scaling $\df_t\tilde{u}= \sqrt{\epsilon}\;\mathcal{U}\; \df_t \tilde{u}'$ is obtained from the variance of the martingale part of the vertical acceleration term \eqref{eq:vmoment2} considering the hydrostatic balance \eqref{eq:hydro1} and the continuity equation \eqref{eq:continu2}. Therefore, for small Rossby number ($\Ro \leq 1$), the random Coriolis term counter-balances the random gradient pressure force:

	\begin{equation}
	\vec{f} \times \sdbth \approx - \frac{1}{\rho_0} \gradh \dpt.
	\end{equation}

Besides, under weak vertical uncertainty, the dimensional continuity equations \eqref{eq:continu2} and \eqref{eq:continu1} reduce to

	\begin{equation}
	\divh \sdbth = \divh \vec{u}_s = 0.
	\end{equation}

As a result, the vertical integration (from bottom topography $\eta_b$ to free surface $\eta$) of the continuity equations \eqref{eq:continu1} become 
\begin{subequations}\label{seq:derive-mass}
	
		\begin{align}
		\left.(w - w_s)\right|_{z = \eta} - \left.(w - w_s)\right|_{z = \eta_b} = - h \divh \vec{u},\ \quad 
		\left.\sigma \df B_t \right|_{z = \eta} - \left.\sigma \df B_t \right|_{z = \eta_b} = 0,
		\end{align}
	
	where $h = \eta - \eta_b$ denotes the thickness of the water column (with a still bottom). On the other hand, a small vertical (Eulerian) displacement at the top and bottom of the fluid leads to a variation of the position of a particular fluid element \cite{Vallis2017}:
	
		\begin{align}
		\left.\big( (w - w_s)\, \dt + \sigma \df B_t \big)\right|_{z = \eta} = \Dfh_t \eta,\ \quad
		\left.\big( (w - w_s)\, \dt + \sigma \df B_t \big)\right|_{z = \eta_b} =  \Dfh_t \eta_b.
		\end{align}
	
\end{subequations}
Combining Equations \eqref{seq:derive-mass}, we deduce the following stochastic mass equation:

	\begin{equation}\label{eq:RSW-mass}
	\Dfh_t h + h \divh \vec{u}\, \dt = 0.
	\end{equation}

Gathering all the elements derived so-far, we finally obtain the following RSW-LU system 

	\begin{subequations}\label{seq:RSWLU}
		\begin{align}
		&{(\text{\em Conservation of momentum})}\nonumber\\
		&\Df_t \vec{u} + \vec{f} \times \vec{u}\, \dt = - g \grad \eta\, \dt, \label{eq:RSWLU-moment}\\
		&{(\text{\em Conservation of mass})}\nonumber\\
		&\Df_t h + h \div \vec{u}\, \dt = 0, \label{eq:RSWLU-mass}\\
		&{(\text{\em  Random balance})}\nonumber\\
		&\vec{f} \times \sdbt = - \frac{1}{\rho} \grad \dpt, \\
		&{(\text{\em  Incompressible constraints})}\nonumber\\
		&\div \sdbt = 0,\ \quad \div \vec{u}_s = 0, \label{eq:RSWLU-incomp}
		\end{align}
	\end{subequations}

where the symbol $H$ for all horizontal variables are dropped  for readability reasons. In ~\ref{A:energy} it is shown that this stochastic system conserves the global energy: 

	\begin{equation} \label{eq:RSWLU-energy} 
	\df_t \int_{\Omega} \frac{\rho}{2} \big( h |\vec{u}|^2 + g h^2 \big)\, \df \vec{x} = 0.
	\end{equation}

It shares thus exactly the same energy conservation property as the deterministic one and beyond their formal resemblance this provides a strong physical link between the two systems. 
Moreover, it can be noticed that under a sufficiently weak (horizontal) uncertainty ($\vec{\sigma} \approx 0$), the system \eqref{seq:RSWLU} reduces to the classical RSW system, in which the stochastic transport operator weighted by the unit of time, $\Df_t / \dt$, reduces to the material derivative.

\section{Structure-preserving discretization of RSW--LU}\label{section-discLU}


In order to perform  numerical simulations of the RSW--LU \eqref{seq:RSWLU} the  noise term $\sdbt$ has to be {\em a priori} parametrized. Its shape is conveniently expressed through a spectral representation and a set of basis functions \eqref{seq:KL}. In this work  homogeneous as well as  heterogeneous spatial structures have been used and the way they are defined  is reviewed in ~\ref{sec:param-noise}.  The incompressible homogenous noise (see Appendix \ref{sec:fft-noise}) is defined through a convolution kernel and is associated with Fourier modes orthogonal functions.  It is easy to implement through fast Fourier transform (FFT). As shown in Section \ref{sec:inviscid}, this noise was in particular used to assess the numerical energy behavior of the discrete scheme. However, homogeneous noises, although carefully scaled from a known energy spectrum established at high resolution, fail to represent inhomogeneity effect encoded by spatially varying variance (the variance is constant and diagonal for homogeneous incompressible noise). This is detrimental to represent large scale effects shaped by the small-scale components in geophysical fluid dynamics. As a matter of fact as shown in \cite{Bauer2020jpo}, heterogeneous noise shapes the large-scale flow in a way akin to the action of vortex force associated with the classical Stokes drift.

In this work, two different parameterizations of heterogeneous noise have been used and are described in Appendix \ref{sec:eof-noise}. The former consists in calibrating empirical orthogonal basis functions (EOF) before the simulation (off-line) from available high-resolution simulation data while the latter consists in specifying the basis functions from the on-going (low resolution) simulation (i.e.  on-line). The second basis functions do not depend on data and are time evolving  whereas the first ones are data driven and stationary. A procedure based on dynamic mode decomposition \cite {Schmid10} to define the noise through evolving basis functions could have been as well used, as proposed by \cite{Gugole2019}. Such a time evolving basis, learned from a high resolution simulation, are shown to perform better that stationary EOF based models. We will have the same type of conclusions for the non-stationnary noise experimented here.
In Section \ref{sec:Galeswky}, both heterogeneous noises are adopted for identifying the barotropic instability of a mid-latitude jet.

In the following, we focus on an energy conserving (in space) approximation of the random dynamical system (RSW--LU).
In this context, the spatial discretization allows us to mimic the balance between the global energy brought by the noise and the LU-diffusion (see Eqn. \ref{eq:energy-balance}) at each time step, hence no additional numerical dissipation or energy increase is introduced into the system.
Considering the definition of 
the stochastic transport operator $\Df_t$ in \eqref{eq:STO}, 
the RSW--LU system in 
Eqn.~\eqref{eq:RSWLU-moment}--\eqref{eq:RSWLU-mass} can be explicitly written as 

	\begin{subequations}
		\begin{align}\label{eq:full_u}
		\df_t \vec{u} = \Big( - \vec{u} \adv \vec{u} - \vec{f} \times \vec{u} - g \grad \eta \Big)\, \dt + \Big( \alf \div \div (\vec{a} \vec{u})\, \dt - \sdbt \adv \vec{u} \Big),
		\end{align}
		\vspace{-1em}
		\begin{align}\label{eq:full_h}
		\df_t h = - \div (\vec{u} h)\, \dt + \Big( \alf \div \div (\vec{a} h)\, \dt - \sdbt \adv h \Big).
		\end{align}
	\end{subequations}

We suggest to develop an approximation of the stochastic RSW--LU model~\eqref{eq:full_u}--\eqref{eq:full_h} by first discretizing the deterministic model underlying this system with a structure-preserving discretization method (that preserves energy in space) and, then, to approximate (with a potentially different discretization method) the stochastic terms. Here, we use for the former a variational discretization approach on a triangular C--grid while for the latter we apply a standard finite difference method. Note that for the methodology introduced in this manuscript,
other spatially energy conserving discretizations rather than the suggested variational integrator 
could be used too. 
The \emph{deterministic dynamical core} of our stochastic system results from simply setting $\vec{\sigma} \approx 0$ in the  equations \eqref{eq:full_u}--\eqref{eq:full_h}. To obtain the full discretized (in space and time) scheme for this stochastic system, we wrap the discrete stochastic terms around the deterministic core and combine this with an Euler--Marayama time scheme. 

Introducing discretizations of the stochastic terms that do not necessarily share the same operators as the deterministic scheme has 
various advantages, as discussed in more detail in Section~\ref{sec_interface}. For instance, such a well defined interface between these two model components minimizes the necessity to adapt the discretization schemes to each other which, in turn, would permit us to apply our method immediately to existing dynamical cores of global numerical weather prediction (NWP) models.

\subsection{Discretization of deterministic RSW equations}\label{section-VarInt}

As mentioned above, the deterministic model (or deterministic dynamical core) of the above stochastic system results from setting $\vec{\sigma} \approx 0$, which 
leads via \eqref{eq:var0} to $\vec{a} \approx 0$. Hence, Equations \eqref{eq:full_u}--\eqref{eq:full_h} reduce to the deterministic RSW equations

	\begin{align}\label{eq:rsw_determ}
	\df_t \vec{u} = \Big( - (\curl{\vec{u}} + \vec{f}) \times \vec{u} -  \grad (\frac{1}{2} \vec{u}^2) - g \grad  \eta  \Big)\,\dt ,
	\qquad
	\df_t h = - \div (\vec{u} h)\, \dt ,
	\end{align}

where we used the vector calculus identity $\vec{u} \adv \vec{u} = (\curl{\vec{u}}) \times \vec{u} + \frac{1}{2} \vec{u}^2$. Note that in the deterministic case  $\df_t / \dt$ agrees (in the limit $\dt\to 0$) with the partial derivative $\partial / \partial t$.

\subsubsection{Variational discretizations}
In the following we present an energy conserving (in space) approximation of these equations 
using a variational discretization approach. While details about the derivation can be found 
in \cite{Bauer2019,Brecht2019}, here we only give the final, fully discrete scheme. 

\begin{figure}
	\centering
	\includegraphics[width=0.7\textwidth]{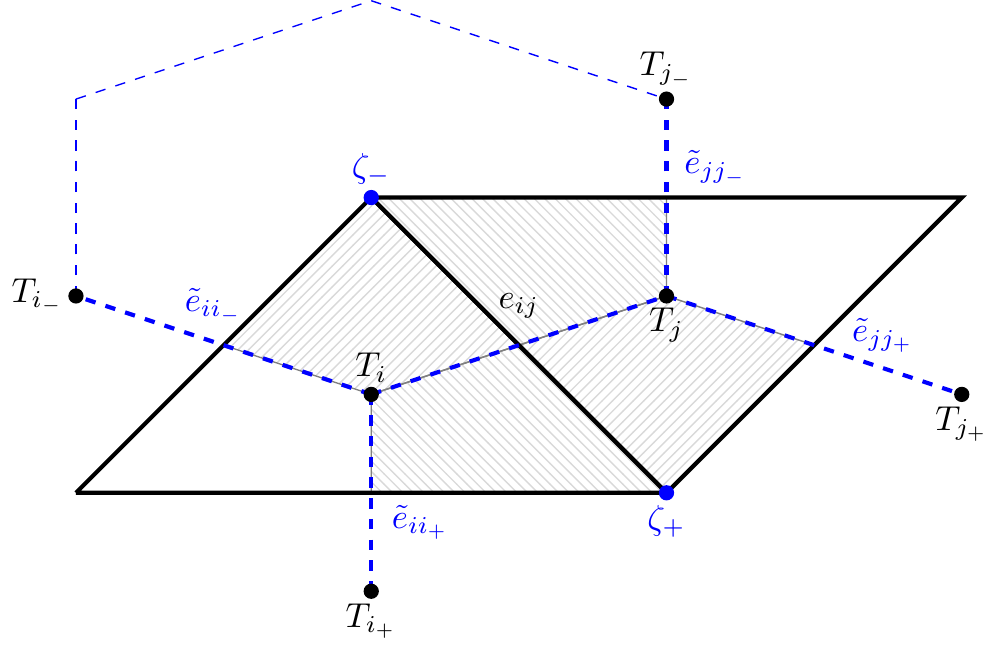}
	\caption{Notation and indexing conventions for the 2D simplicial mesh.}
	\label{fig:Notationgrid}
\end{figure}

To do so, we start with introducing the mesh and some notation. 
The variational discretization of \eqref{eq:rsw_determ} results in a scheme that corresponds to a C-grid staggering of the variables on a quasi uniform triangular grid with hexagonal/pentagonal dual mesh. 
Let $N$ denote the number of triangles used to discretize the domain. 
As shown in Fig. \ref{fig:Notationgrid}, we use the following notation: $T$ denotes the primal triangle, $\zeta$ the dual hexagon/pentagon,  
$e_{ij}=T_i\cap T_j$ the primal edge and
$\tilde{e}_{ij}=\zeta_{+}\cap \zeta_{-}$ the associated dual edge. Furthermore, we have $\mathbf{n}_{ij}$ and $\mathbf{t}_{ij}$ as the normalized normal and tangential vector relative to edge $e_{ij}$ at its midpoint.
Moreover, $D_i$ is the discrete water depth at the circumcentre of $T_i$, 
${\eta_b}_i$ the discrete bottom topography at the circumcentre of $T_i$, 
and $V_{ij}=(\mathbf{u}\cdot \mathbf{n})_{ij}$ the normal velocity at the triangle edge midpoints in the direction from triangle $T_i$ to $T_j$. We denote $\overline{D}_{ij}=\frac{1}{2}(D_i+D_j)$ as the water depth averaged to the edge midpoints. 

The variational discretization method does not require to define explicitly approximations of 
the differential operators because they directly result from the discrete variational principle. 
It turns out that on the given mesh, these operators agree with the following definitions of 
standard finite difference and finite volume operators:

	\begin{equation}
	\begin{aligned}[l]							
	(\text{Grad}_n~F)_{ij}&\defin\frac{F_{T_j}-F_{T_i}}{|\tilde{e}_{ij}|},
	\\
	(\text{Grad}_t~F)_{ij}&\defin\frac{F_{\zeta_-}-F_{\zeta_+}}{|e_{ij}|},
	\end{aligned}	\qquad \qquad
	\begin{aligned}[l]
	(\text{Div}~V)_i&\defin\frac{1}{|T_i|}\sum_{k\in\{j,i_-,i_+\}}|e_{ik}|V_{ik},
	\\
	(\text{Curl}~V)_{\zeta}&\defin\frac{1}{|\zeta|}\sum_{\tilde{e}_{nm}\in\partial\zeta} |\tilde{e}_{nm}|V_{nm},
	\end{aligned}
	\end{equation}

for the normal velocity $V_{ij}$ and a scalar function $F$ either sampled as $F_{T_i}$ at the circumcentre of the triangle $T_i$ or sampled as $F_{\zeta_{\pm}}$  at the centre of the dual cell $\zeta_{\pm}$.
The operators $\text{Grad}_n$ and $\text{Grad}_t$ correspond to the gradient in the normal and tangential direction, respectively,
and $\text{Div}$ to the divergence of a vector field:

	\begin{align}
	(\nabla F)_{ij} &\approx (\text{Grad}_n~F)\mathbf{n}_{ij}+ (\text{Grad}_t~F)\mathbf{t}_{ij}, \label{eq:discreteGrad}
	\\
	(\nabla\cdot \mathbf{u})_{i}&\approx (\text{Div}~V)_i,
	\\
	(\nabla\times \mathbf{u})_{\zeta}&\approx (\text{Curl}~V)_\zeta. \label{eq:vorticity}
	\end{align}

The last equation defines the discrete vorticity and for later use, we also discretize the potential vorticity as 
 
	\begin{equation}\label{eq:pv}
	\frac{\nabla\times \mathbf{u}+f}{h} \approx \frac{(\text{Curl}~V)_\zeta+f_\zeta}{D_\zeta}, \qquad \qquad D_\zeta = \sum_{\tilde{e}_{ij}\in\partial \zeta} \frac{|\zeta\cap T_i|}{|\zeta|} D_i. 
	\end{equation}

\subsubsection{Semi-discrete RSW scheme}
With the above notation, the deterministic semi-discrete RSW equations read:

	\begin{subequations}
		\begin{equation}\label{eq:mom}
		\df_t V_{ij} = \mathcal{L}_{ij}^{\sub V}(V,D)\, \Delta t, \quad \text{for all edges } e_{ij},
		\end{equation}
		\vspace{-1em}
		\begin{equation}\label{eq:cont}
		\df_t D_{i} = \mathcal{L}_{i}^{\sub D}(V,D)\, \Delta t, \quad \text{for all cells } T_i,
		\end{equation}
	\end{subequations}

where $\mathcal{L}_{ij}^{\sub V}$ and $\mathcal{L}_{i}^{\sub D}$ denote the deterministic spatial operators, and $\Delta t$ stands for the discrete time step. 
The RHS of the momentum equation~\eqref{eq:mom} is given by

	\begin{equation}
	\mathcal{L}_{ij}^{\sub V}(V,D) \defin -\text{Adv}(V,D)_{ij} - \text{K}(V)_{ij}-\text{G}(D)_{ij} ,
	\end{equation}

where Adv denotes the discretization of the advection term  $(\curl{\vec{u}} + \vec{f}) \times \vec{u}$ of \eqref{eq:rsw_determ}, 
K the approximation of the gradient of the kinetic energy $\grad (\frac{1}{2} \vec{u}^2)$ and G of the gradient of the height field 
$g \grad \eta$. Explicitly, the advection term is given by

	\begin{equation}
	\begin{split}
	& \operatorname{Adv}_{\rm }(V,D)_{ij} \defin \\ 
	&   -  \frac{1}{\overline{D}_{ij} |\tilde{e}_{ij}| } \Big((\text{Curl } V)_{\zeta_{-}}+f_{\zeta_{-}}\Big)
	\left(   \frac{|\zeta_{-} \cap T_i|}{2 |T_i|  } \overline{D}_{ji_-} |e_{ii_-}|   V_{ii_-} 
	+  \frac{|\zeta_{-}  \cap T_j|}{2 |T_j|  } \overline{D}_{ij_-}  |e_{jj_-}|   V_{jj_-} \right) \\
	& +  \frac{1}{\overline{D}_{ij}  |\tilde{e}_{ij}| }\Big((\text{Curl } V)_{\zeta_{+}}+f_{\zeta_{+}}\Big) \left(  \frac{|\zeta_{+}  \cap T_i|}{2 |T_i| }\overline{D}_{ji_+} |e_{ii_+}|   V_{ii_+}
	+ \frac{|\zeta_{+}  \cap T_j|}{2 |T_j|   } \overline{D}_{ij_+}   |e_{jj_+}|   V_{jj_+} \right),
	\end{split}
	\end{equation}

where $f_{\zeta_{\pm}}$ is the Coriolis term evaluated at the centre of $\zeta_{\pm}$.
Moreover, the two gradient terms read:

	\begin{align} 
	& \operatorname{K}_{\rm }( V) _{ij} \defin \frac{1}{2}(\text{Grad}_n ~F)_{ij}, 
	\qquad\qquad  F_{T_i}= \sum_{k\in \{j, i_-, i_+\} }\frac{|\tilde{e}_{ik}|~|e_{ik}|(V_{ik})^2}{2|T_k|},
	\\
	&\operatorname{G}(D)_{ij} \defin  g (\text{Grad}_n~(D+\eta_b))_{ij}. 
	\end{align}

The RHS of the continuity equation ~\eqref{eq:cont} is given by 

	\begin{equation}	
	\mathcal{L}_{i}^{\sub D}(V,D) \defin -\big(\text{Div}~(\overline{D}V)\big)_{i},
	\end{equation}

which approximates the divergence term 
$-\div (\vec{u} h)$. 


\subsubsection{Time scheme}
For the time integrator we use a Crank-Nicolson-type scheme where we solve the system of fully discretized non-linear momentum and continuity equations by a fixed-point iterative method. The corresponding algorithm coincides for $\vec{\sigma}=0$ with the one given in Section \ref{sec:EulerMaru}.

\subsection{Spatial discretization of RSW--LU}\label{section-stochscheme}
The fully stochastic system has additional terms on the RHS of Equations \eqref{eq:full_u} and \eqref{eq:full_h}. With these terms the discrete equations read:

	\begin{subequations}
		\begin{equation}
		\df_t V_{ij} = \mathcal{L}_{ij}^{\sub V}(V,D)\, \Delta t + \Delta \mathcal{G}_{ij}^{\sub V},
		\end{equation}
		\vspace{-1em}
		\begin{equation}
		\df_t D_{i} = \mathcal{L}_{i}^{\sub D}(V,D)\, \Delta t + \Delta \mathcal{G}_{i}^{\sub D},
		\end{equation}
		where the stochastic LU-terms are given by 
		\begin{equation}\label{eq:discreteMomentumLU}
		\Delta \mathcal{G}_{ij}^{\sub V} \defin \Big(  \frac{\Delta t}{2}\, \big(\boldsymbol{\nabla\cdot\nabla\cdot}    (\vec{a} \vec{u})\big)_{ij}  - (\sdbt \adv \vec{u})_{ij} \Big) \bdot \vec{n}_{ij},
		\end{equation}
		\vspace{-1em}
		\begin{equation}\label{eq:discreteMassLU}
		\Delta \mathcal{G}_{i}^{\sub D} \defin  \frac{\Delta t}{2} \big( \boldsymbol{\nabla\cdot\nabla\cdot}  (\vec{a} D)\big)_{i}  - (\sdbt \adv D)_{i}.
		\end{equation}
	\end{subequations}

Note that the two terms within the large bracket in \eqref{eq:discreteMomentumLU} 
comprise two Cartesian components of a vector 
which is then projected onto the triangle edge's normal direction via $\vec{n}_{ij}$.
The two terms in \eqref{eq:discreteMassLU} are scalar valued at the cell circumcenters $i$.

The parametrization of the noise described in \ref{sec:param-noise} is
formulated in Cartesian coordinates, because this allows using standard algorithms to 
calculate EOFs, for instance. 
Likewise, we represent the stochastic LU-terms in Cartesian coordinates but to connect both 
deterministic and stochastic terms, 
we will calculate the occurring differentials with operators as provided by the 
deterministic dynamical core (see interface description below).
Therefore, we write the second term in \eqref{eq:discreteMomentumLU} as

	\begin{equation}	\label{eq:secondtermLU}
	(\sdbt \adv F)_{ij}= \sum_{l=1}^2 (\sdbt)_{ij}^l (\nabla F)_{ij}^l,
	\end{equation}

in which ${(\sdbt)}_{ij}$ denotes the discrete noise vector with two Cartesian components, constructed as described in \ref{sec:param-noise}
and evaluated at the edge midpoint $ij$. 
The scalar function $F$ is a placeholder for the Cartesian components of the velocity field $\vec{u} = (u^1,u^2)$.
Likewise, the first term in \eqref{eq:discreteMomentumLU} can be written component-wise as 

	\begin{equation}\label{eq:rewriteDiffu}
	(\boldsymbol{\nabla\cdot\nabla\cdot} (\vec{a}F))_{ij}
	=\sum_{k,l=1}^2 \left(\partial_{x_k} \left( \partial_{x_l} (a_{kl}F)\right)_{ij} \right)_{ij}
	,
	\end{equation}

where $a_{kl}$ denotes the matrix elements of the variance tensor which will be 
evaluated, similarly to the discrete noise vector, at the edge midpoints. 
For a concrete realization of the differentials on the RHS of both stochastic terms,
we will use the gradient operator \eqref{eq:discreteGrad} as introduced next.

To calculate the terms in \eqref{eq:discreteMassLU} we also use the representations \eqref{eq:secondtermLU} 
and \eqref{eq:rewriteDiffu} for a scalar function $F=D$ describing the water depth. However, as our proposed procedure will 
result in terms at the edge midpoint $ij$, we have to average them to the cell centers $i$.

In the following, we will refer to this part of the code that 
generates the noise on a Cartesian mesh according to \ref{sec:param-noise} 
as \emph{noise generation module.}

\subsubsection{Interface between dynamical core and LU terms}\label{sec_interface}


As mentioned above, the construction of the noise is done on a Cartesian mesh while the 
discretization of the deterministic dynamical core (variational RSW scheme, Section~\eqref{section-VarInt}), 
corresponding to a triangular C-grid staggering, 
predicts the values for velocity normal to the triangle edges and for water depth at the triangle centers. 
We propose to exchange information between the noise generation module (see section above) and the dynamical core 
via the midpoints of the triangle edges where on such C-grid staggered discretizations 
the velocity values naturally reside. The technical details about how we realized such 
interface in our setup are given in \ref{A:LUdiscrete}.

This modular approach with a well defined interface between these two model 
components has various advantages over directly implementing the noise terms 
on a triangular C-grid mesh as used by the dynamical core. 
Firstly, this approach allows us to easily explore various noise types, because 
using a Cartesian mesh for the latter permits the usage of standard algorithms for e.g. FFT or singular value decomposition (SVD).
In contrast, exploring these ideas directly on a triangular C-grid would significantly increase 
the implementation work. In fact, this manuscript also serves as a proof of 
concept study to show that such modular approach indeed works very well.

Moreover, the definition of an interface between the two model components 
should minimize (or maybe even avoid) the necessity of adapting the numerics of an existing 
deterministic core in order to incorporate the discrete stochastic LU-terms. 
This, in turn, should allow us to apply our method directly to existing dynamical cores of NWP 
models. 

\subsubsection{Computational aspects} 
In addition to the deterministic scheme we have the terms $ \Delta \mathcal{G}^{\sub V}$ and  $\Delta \mathcal{G}^{\sub D}$ for the RSW--LU scheme (see Eq. \eqref{eq:discreteMomentumLU} and Eq. \eqref{eq:discreteMassLU}). Their discretization can be differentiated into:
\begin{itemize}
	\item \underline{The noise generation of $\sdbt$ and $\vec{a}$}.
	The noise generation relies on generating a fixed number of pseudo-observations and carrying out a SVD to obtain the EOFs. The SVD can be carried out as an economy-size SVD which depends linearly on the number of triangles.
	Currently for LU on-line, EOFs are estimated at each time step, but less frequent estimations are also possible to save computational costs.
	
	\item \underline{The computation of the divergence  and gradient in Cartesian coordinates}. The discretization of these operations are described in \ref{A:LUdiscrete}, which results in matrix vector multiplications.
\end{itemize}

Here, we obtain  the discretization of $ \Delta \mathcal{G}^{\sub V}$ and  $\Delta \mathcal{G}^{\sub D}$ using the interface, which is determined by the underlying discretization of the deterministic scheme. More specifically, we reformulate the differential operators in Cartesian coordinates with the local derivatives obtained from the deterministic scheme (see e.g. Eq. \eqref{eq:discretePartial}). 
This results only in a few additional matrix vector multiplications. 

Optimized standard methods for the noise generation on a Cartesian mesh are potentially 
more efficient than a direct (and not optimized) implementation on a triangular mesh.
Besides the advantages mentioned above and given that the additional computational costs 
for interchanging the values via the interface consists of only a few matrix vector 
multiplications, we advocate our modular approach rather than a direct implementation.

\subsection{Temporal discretization of RSW--LU} \label{sec:EulerMaru}

The iterated Crank-Nicolson method presented in \cite{Brecht2019} is adopted for the temporal discretization.
Keeping the iterative solver and adding the LU terms results in an
Euler-Maruyama scheme, which decrease the order of convergence of the deterministic iterative solver (see \cite{Kloeden1992} for details).

To enhance readability, we denote $V^t$ as the array over all edges $e_{ij}$ of the velocity $V_{ij}$ and $D^t$ as the array over all cells $T_i$ of the water depth $D_i$ at time $t$. The governing algorithm reads:

\medskip

\begin{algorithm}[H] \label{sec:RSWLUalg}
	\SetAlgoLined
	
	Set iterative solver index $k=0$ with initial guess at  $t$:
	
		\begin{align*}
		V^*_{k=0}&=V^t,  
		\\
		(D_{k=0}^*)&=D^t+ \Delta \mathcal{G}^{\sub D}(D^t),
		\end{align*}
	
	and compute $\Delta \mathcal{G}_{ij}^{\sub V}(V^t)$.
	
	\While{$\|V^*_{k+1}-V^*_{k}\|+\|D^*_{k+1}-D^*_{k}\|>\text{ tolerance}$}{
		
			\begin{align*}
			\frac{D^*_{k+1}-D^t}{\Delta t}=&-\frac{\text{Div}~(\overline{D^*_k}V^*_k)+\text{Div}~(\overline{D^t}V^t)}{2}
			\\
			\frac{V^*_{k+1}-V^t}{\Delta t} =&  -\frac{\operatorname{Adv}_{\rm }(V^*_k,D^*_{k+1})+\operatorname{Adv}_{\rm }(V^t,D^t)}{2}
			-\frac{\operatorname{K}_{\rm }(V^*_k)+\operatorname{K}_{\rm }(V^t)}{2}
			-\operatorname{G}_{\rm }(D^*_{k+1}) \\
			& 
			+ \Delta \mathcal{G}_{ij}^{\sub V}(V^t) 
			\end{align*}
		
		and set $k+1 = k$.
		
	}
	\caption{Time-stepping algorithm}
\end{algorithm}

For all simulations in this manuscript, we used a tolerance of $10^{-6}$ for simulations on the f-plane and $10^{-10}$ for simulation on the sphere. 
In all these cases, our suggested fixed point solver converges in less than 10 iterations.

\section{Numerical results}\label{sec:results}

In this section, we first study the energy behaviour of the numerical RSW--LU scheme introduced above for an inviscid test flow. Then, we show that for a viscous test case, the stochastic model captures more accurately the reference structure of the large-scale flow when compared to the deterministic model under the same coarse resolution. In addition, we demonstrate that the proposed RSW--LU system provides a more reliable ensemble forecast with larger spread, compared to a classical random model based on the perturbations of the initial conditions (PIC).


\subsection{Inviscid test case~--~energy analysis}\label{sec:inviscid}

This first test case consists of two co-rotating vortices on the $f$-plane. To illustrate the energy conservation of the spatial discretization of the RSW--LU system \eqref{seq:RSWLU}, we use the homogeneous stationary noise defined in Section \ref{sec:fft-noise} since the two incompressible constraints $\div \sdbt = 0$ and $\div \div \vec{a} = 0$ in \eqref{eq:RSWLU-incomp} are naturally satisfied.
Then, no extra steps are required to satisfy the incompressible constraints.

\subsubsection*{Initial conditions}

The simulations are performed on a rectangular double periodic domain $\Omega = [0, L_x] \times [0, L_y]$ with $L_x = 5000\, \text{km}$ and $L_y = 4330\, \text{km}$, which is discretized into $N=32768$ triangles. 
We use this resolution for both the deterministic and stochastic simulations. 
The large-scale flow is assumed to be under a geostrophic regime at the initial state, {\em i.e.} $f \vec{k} \times \vec{u} = - g \grad h$. 
We use an initial height field elevation (as e.g. in \cite{Bauer2019}) of the form
\begin{subequations}
	
		\begin{equation}
		h \big( x, y, t=0 \big) = H_0 - H' \bigg( \exp \Big( -\frac{{x'_1}^2+{y'_1}^2}{2} \Big) + \exp \Big(-\frac{{x'_2}^2+{y'_2}^2}{2} \Big) -\frac{4\pi s_x s_y}{L_x L_y} \bigg),
		\end{equation}
	
	where the background height $H_0$ is set to $10\, \text{km}$, the magnitude of the small perturbed height $H'$ is set to $75\, \text{m}$ and the periodic extensions $x'_i, y'_i$ are given by
	
		\begin{equation}
		x'_i = \frac{L_x}{\pi s_x} \sin \big( \frac{\pi}{L_x} (x - x_{c_i}) \big),\ \quad 
		y'_i = \frac{L_y}{\pi s_y} \sin \big( \frac{\pi}{L_y} (y - y_{c_i}) \big),\ \quad
		i = 1,2
		\end{equation}
	
	with the centres of the vertices located at $(x_{c_1}, y_{c_1}) = \frac{2}{5}\, (L_x, L_y)$, $(x_{c_2}, y_{c_2}) = \frac{3}{5}(L_x, L_y)$ with parameters $(s_x , s_y) = \frac{3}{40} \, (L_x, L_y)$.
\end{subequations}
To obtain the discrete initial water depth $D_i$, we sample the analytical function $h$ at each cell centre. Subsequently, the discrete geostrophic velocities 
at each triangle edge $ij$  at the initial state can be deduced via

	\begin{equation}
	V_{ij} = -\frac{g}{f} (\text{Grad}_t~D)_{ij},\ 
	\end{equation}

where the Coriolis parameter $f$ is set to $5.3108 \text{ days} ^{-1}$. For the LU simulations, the magnitude of the homogeneous noise remains moderate with its constant variance $a_0$ set to be $169.1401\, \text{m}^2 \cdot \text{s}^{-1}$.

\begin{figure}
	\includegraphics[trim=3cm 3cm 1cm 4cm, clip,width=\textwidth]{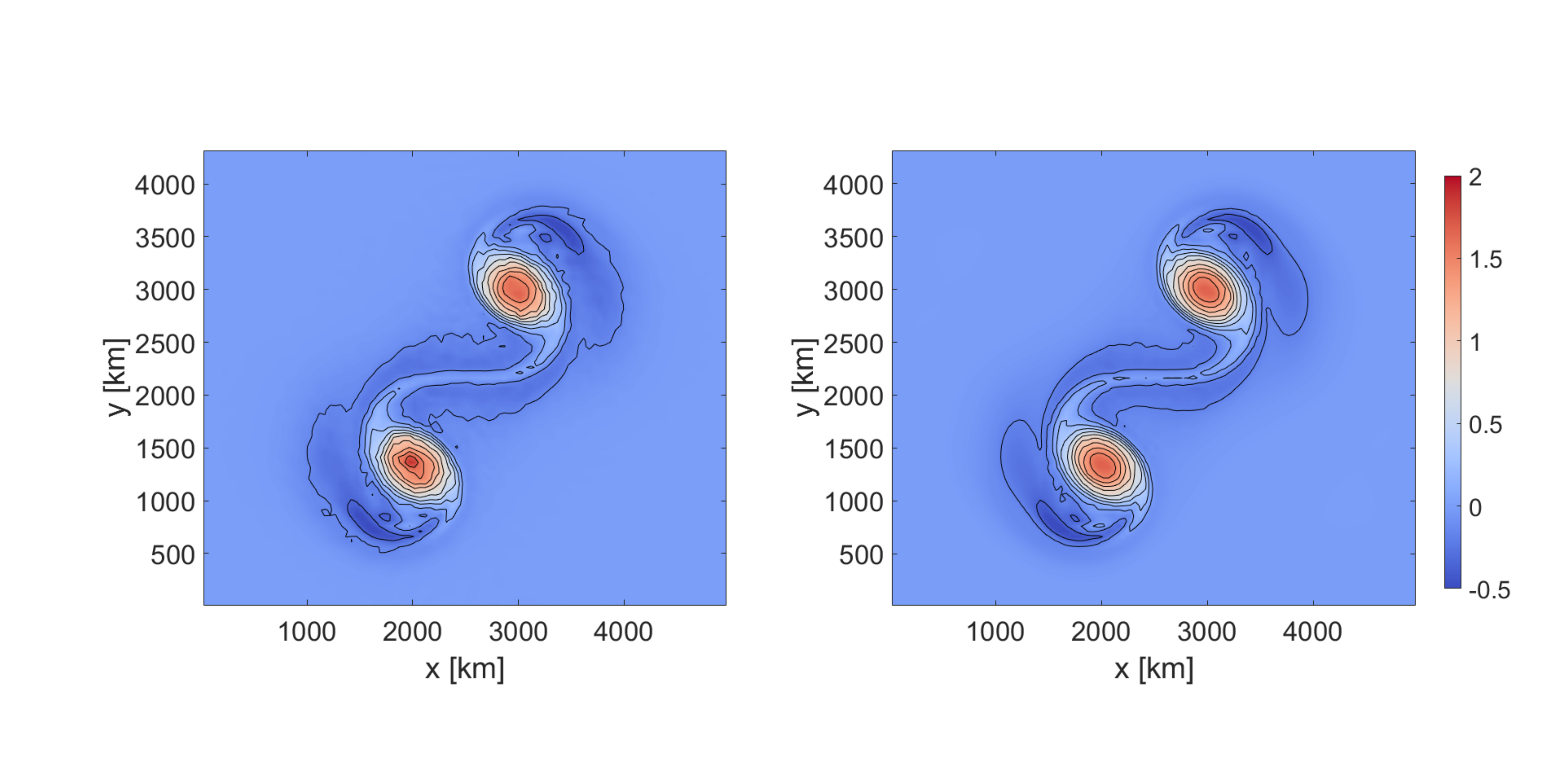}
	\caption{Contour plots of the potential vorticity fields after 2 days for (left) one realization of a LU simulation with homogeneous noise and (right) a deterministic run. The contour interval is 0.4 $\text{days}^{-1}\text{ km}^{-1}$.}
	\label{fig:luVsOpsDvtx}
\end{figure}

\subsubsection*{Analysis of energy conservation}

To analyze the energy conservation properties of our stochastic integrator, we use the above initial conditions to simulate the two co-rotating vortices for 2 days. In Figure \ref{fig:luVsOpsDvtx}, we show contour plots of the potential vorticity (as defined in \eqref{eq:pv}) fields of the deterministic and stochastic models. We observe that under the moderate noise with $a_0$ as chosen above, the large-scale structure of the stochastic system is similar to that of the deterministic run.

On the specific staggered grid as shown in Figure \ref{fig:Notationgrid},
the total energy of the shallow water equations \eqref{seq:E-RSW}, for both deterministic and stochastic case, 
is approximated by

	\begin{equation}\label{e_tot}
	\E (t) \approx \sum_{i=1}^N \frac{1}{2} D_{i}(t) |T_i| \sum_{k=j, i_-, i_+} \frac{1}{2|T_i|} h_{ik}f_{ik} \big( V_{ik} (t) \big)^2 + \frac{1}{2} g \big( D_{i} (t) \big)^2 |T_i|.
	\end{equation}

As shown in \cite{Bauer2019}, the proposed discrete variational integrator (see Section~\ref{section-VarInt}) together with
an iterative Crank-Nicolson time stepping 
method exhibits a 1st order convergence rate of the energy error with smaller time step size.
This will allows us immediately to simply include the stochastic terms to result  
in an Euler-Maruyama type time integrator for stochastic systems (cf. Section \ref{section-stochscheme}).

\begin{figure}
	\begin{center}
		\includegraphics[width=0.8\textwidth]{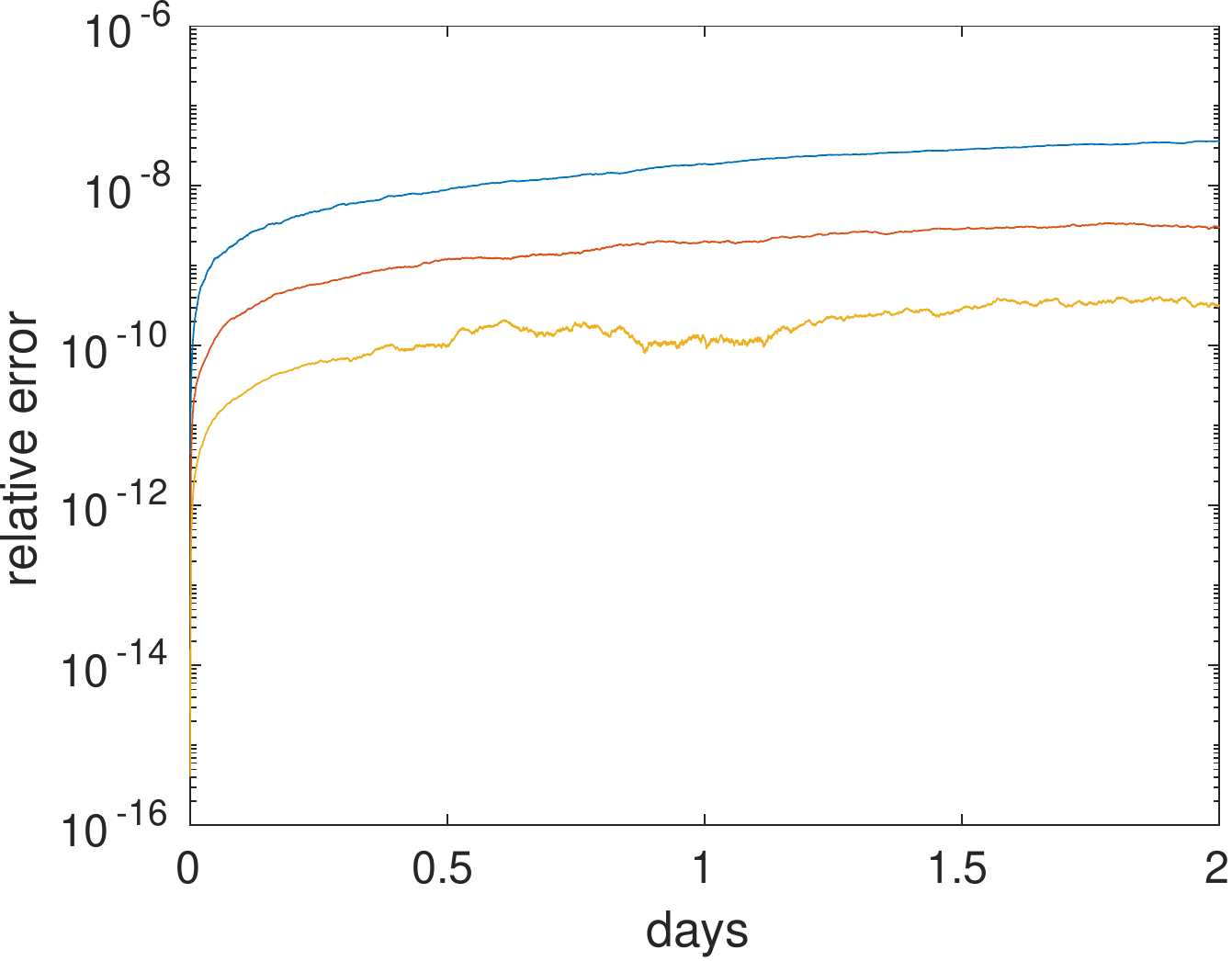}
	\end{center}
	\caption{Evolution of the relative $L_2$ errors between the energy of the mean RSW--LU and the reference, using $\Delta t$ (blue line), $\Delta t / 10$ (red line) and $\Delta t /100$ (yellow line) respectively.} \label{fig:enConv}
\end{figure}

In the present work, we consider the energy behavior of the deterministic scheme (i.e. the variational integrator) as reference, which is denoted as $\E_{\sub\text{REF}} (t)$ in the following. For the stochastic RSW model, the Euler-Maruyama time scheme might lead to a different behavior with respect to energy conservation when compared to the deterministic model. In order to quantify numerically the energy conservation of the RSW--LU, we propose to measure the relative errors between the mean stochastic energy, denoted as $\overline{\E}_{\sub\text{LU}} (t)$, and the reference $\E_{\sub\text{REF}} (t)$ by $ \overline{\E}_{\sub\text{LU}} (t) / \E_{\sub\text{REF}} (t) - 1$, while using for both the same spatial resolution (see Table~\ref{tab:params}). This setup allows us to measure the influence of the stochastic terms on the energy conservation relative to the deterministic scheme. Figure \ref{fig:enConv} shows these relative errors for different time step sizes over a simulation time of 2 days. As we can confirm from the curves, taking successively smaller time steps\\ $\Delta t\in\{ 1.7361 \times 10^{-4}, 
3.4722 \times 10^{-5}, 
1.7361 \times 10^{-5}, 
3.4722 \times 10^{-6}, 
1.7361 \times 10^{-6} \}$ (in $\text{days}^{-1}$) results in smaller relative errors.

\begin{figure}
	\begin{center}
		\includegraphics[width=0.8\textwidth]{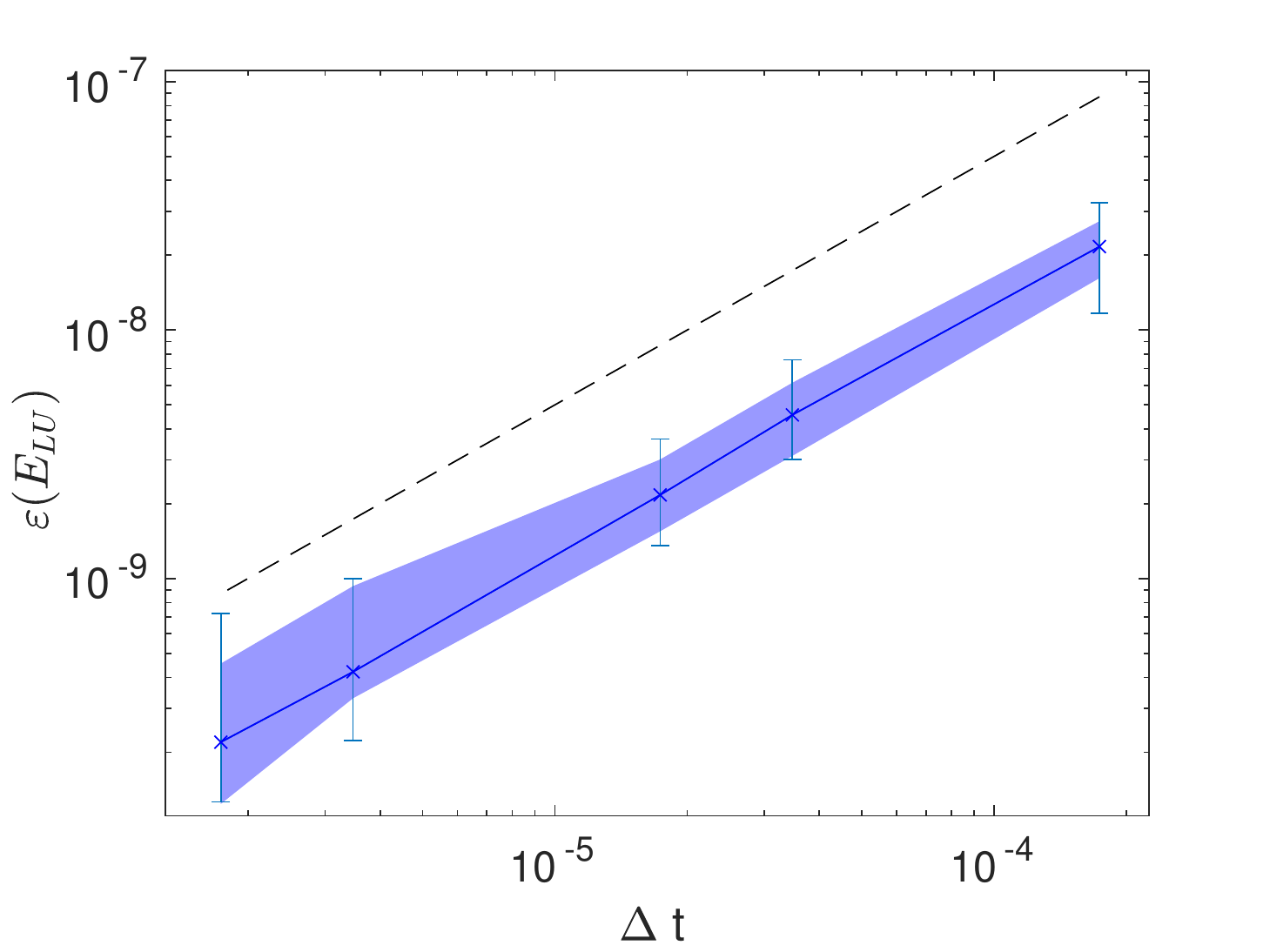}
	\end{center}
	\caption{Convergence of the energy path of the RSW--LU to that of the reference w.r.t. time step sizes. The blue line shows the global errors of the ensemble mean energy, the blue area describes the $68\%$ confident interval of the ensemble errors and the dashed line stands for the 1st order convergence rate. }\label{fig:enConv1stord}
\end{figure}

To determine more quantitatively the convergence rate of the stochastic scheme (relative to the reference) with respect to different time step sizes, we defined the following global (in space and time) error measure:

	\begin{equation}
	\varepsilon (\E_{\sub\text{LU}}) \defin \frac{ \| \E_{\sub\text{LU}} (t) - \E_{\sub\text{REF}} (t) \|_{\sub L^2([0,T])} }{ \| \E_{\sub\text{REF}} (t) \|_{\sub L^2([0,T])} },
	\end{equation}

where $\| f(t) \|_{\sub L^2 ([0,T])} = ( \int_{\sub 0}^{\sub T} |f(t)|^2 \dt )^{1/2}$ and $T$ is set to 2 days. 
We determine for an ensemble with 10 members such global errors in order to illustrate the convergence rate of each ensemble member and the spread between those rates. This spread is illustrated as blue shaded area in Figure \ref{fig:enConv1stord}. 
The area centre is determined by the mean of the errors, and the dispersion of this area is given by one standard derivation 
(\emph{i.e.} $68\%$ confident interval of the ensemble of $\varepsilon (\E_{\sub\text{LU}})$). Besides, the minimal and maximal values of 
the errors of the ensemble are represented by the vertical bar-plots. The blue line of Figure \ref{fig:enConv1stord} shows that the convergence rate (w.r.t. various $\Delta t$) of the ensemble mean energy is of 1st order. This is consistent with the weak convergence rate of order $\mathcal{O} (\Delta t)$ of the Euler-Maruyama scheme, cf. Section~\ref{sec:EulerMaru}.

\subsection{Viscous test case~-~ensemble prediction}\label{sec:Galeswky}


Next, we want to show that our stochastic system better captures the structure of a large-scale flow than a comparable deterministic model.
To this end, we use a viscous test case and heterogeneous noise.

The viscous test case we use is proposed by \cite{Galewsky2004} and it consists of a barotropically unstable jet at the mid-latitude on the sphere. This strongly non-linear flow will be destabilized by a small perturbation of the initial field, which induces decaying turbulence after a few days. However, the development of the barotropic instability in numerical simulations highly depends on accurately resolving the small-scale flow, which is particularly challenging for coarse-grid simulations. For the same reason, the performance of an ensemble forecast system in this test case is quite sensible to the numerical resolution. In the following, we demonstrate that the RSW--LU simulation on a coarse mesh under heterogeneous noises, provides better prediction of the barotropic instability compared to the deterministic coarse simulation, and produces more reliable ensemble spread than the classical PIC simulation.

\subsubsection*{Stabilization}

The former test case \ref{sec:inviscid} consists of smooth enough fields such that no additional sub-grid dissipation is required. In contrast, the following test case consists of the evolution of decaying turbulence, in which sub-grid enstrophy will accumulate quickly, hence an efficient dissipation mechanism is needed, such as the biharmonic  eddy viscosity \cite{Galewsky2004} which is often used in atmospheric and oceanic flow models. Here, we include a biharmonic eddy viscosity with uniform coefficient $\mu$ 
(of unit $m^4/s$) in the momentum equation:

	\begin{align}
	\df_t V &= \Big( -\text{Adv}(V,D)_{ij}-\text{K}(V)_{ij}-\text{G}(D)_{ij}-\mu L(V)_{ij} \Big)\, \Delta t,
	\label{eq:discreteSWE_bilap}
	\end{align}

where:

	\begin{equation}\label{eq:lap}
	L(V)_{ij}=\big( \text{Grad}_n(\text{Div} ~V)_{ij}-\text{Grad}_t(\text{Curl}~V)_{ij} \big)^2.
	\end{equation}


Although in the evolution equation~\eqref{eq:full_u} 
the dissipative term is energetically exactly in balance with the random advection term, 
the supplementary biharmonic diffusion is needed here in this test case to drain the enstrophy pile-up. Using instead a dissipative discretization, in which numerical diffusion takes the role of such stabilization, might give stable simulations also without explicit diffusion but then we would lose control of the strength of the diffusion. Note that we used standard biharmonic dissipation, but there exist also energy conserving enstrophy dissipation methods, such as those introduced in \cite{mcrae2014energy} or in \cite{Frank2003}.

\subsubsection*{Initial conditions}

The values of the principle parameters for the simulations are specified in Table \ref{tab:params}. Under the geostrophic regime, the initial zonal velocity and height is respectively given by
\begin{subequations}\label{seq:init-Galeswky}
	
		\begin{equation}
		u (\Theta, t=0) = \frac{U_0}{e_n} \exp \Big( \frac{1}{(\Theta - \Theta_0)(\Theta - \Theta_1)} \Big),\ \quad \text{for}\ \Theta_0 < \Theta < \Theta_1,
		\end{equation}
		\begin{equation}
		h (\Theta, t=0) = H_0 - \frac{R}{g} \int_{\Theta} u (\theta, t=0) \Big( 2 \tilde{\Omega} \sin \theta + \frac{\tan \theta}{R} u(\theta, t=0) \Big)\, \df \theta,
		\end{equation}
	
	where $e_n = \exp \big( -4 / (\Theta_1 - \Theta_0)^2 \big)$ is used to rescale the jet magnitude to the maximal value $U_0$ at the jet's mid-point $\Theta = \pi / 4$. As introduced by \cite{Galewsky2004}, in order to initiate the barotropic instability, the following localized bump is included in the height field:
	
		\begin{equation}
		h' (\Upsilon, \Theta) = H' \cos \Theta\, \exp \Big( - (3 \Upsilon)^2 - \big( 15 (\frac{\pi}{4} - \Theta) \big)^2 \Big),
		\end{equation}
	
\end{subequations}
where $\Upsilon$ denotes the longitude. 
Here, the Coriolis parameter is set to $f=2 \times 7.292 \times 10^{-5} \sin(\Theta)$.
Analogously to the previous inviscid test case, we then use these analytic functions \eqref{seq:init-Galeswky} to sample the discrete velocity at the edge mid-points and the height field at the cell centres on the staggered mesh (See Figure \ref{fig:Notationgrid}).

\begin{table}[htbp]
	\begin{center}
		\begin{tabular}{ccc}
			\hline
			Parameters             & Value                                             & Description                      \\ \hline
			$(\Theta_0, \Theta_1)$ & $(2\pi, 5\pi) / 14\, \text{rad}$                            & Initial latitude limits          \\
			$H_0$                  & $10.158\, \text{km}$                              & Background height                \\
			$H'$                   & $120\, \text{m}$                                  & Initial perturbation amplitude   \\
			$R$                    & $6.371 \times 10^3\, \text{km}$                   & Mean radius of Earth             \\
			$g$                    & $9.806\, \text{m} \cdot \text{s}^{-2}$                         & Gravity of Earth                 \\
			$\tilde{\Omega}$       & $7.292 \times 10^{-5}\, \text{s}^{-1}$     & Angular rotation rate of Earth   \\
			$U_0$                  & $80\, \text{m} \cdot \text{s}^{-1}$                            & Maximum zonal velocity       \\
			$\mu_{\sub l}$                & $3.975 \times 10^{14}\, \text{m}^4 \cdot \text{s}^{-1}$ & Fine-grid biharmonic viscosity   \\
			$\mu_{\sub L}$                & $3.199 \times 10^{16}\, \text{m}^4 \cdot \text{s}^{-1}$ & Coarse-grid biharmonic viscosity \\ 
			$\Delta t_{\sub l}$                  & $12\, \text{s}$                              & Fine-grid time step               \\
			$\Delta t_{\sub L}$                  & $50\, \text{s}$                              & Coarse-grid time step               \\
			$N_{\sub l}$ & $327680$ & Number of triangles for fine grid (60-km resolution)\\
			$N_{\sub L}$ & $20480$ & Number of triangles for coarse grid (240-km resolution)\\
			\hline
		\end{tabular}
		\caption{Parameter list for simulations of the barotropic instability.}
		\label{tab:params}
	\end{center}
\end{table}

For the LU simulations, we use the two heterogeneous noises described in Section \ref{sec:eof-noise}, based on either the off-line learning of EOFs from the high-resolution simulation data, denoted as LU off-line, or on the on-line estimation of EOFs from the coarse-grid simulation, denoted as \emph{LU on-line}. To allow for comparisons, the strength of these two noises are imposed to be the same. 
The PIC stochastic model is obtained as follows: first, we perform ensemble simulations of the LU off-line and the LU on-line method over 1 day. Then, each realization of these ensemble runs is used as one initial random state and simulated for the remaining days using the deterministic scheme. We call the PIC simulation using the LU off-line method  \emph{PIC 1} and the PIC simulation obtained using the LU on-line method \emph{PIC 2}. 
For each of these stochastic models, an ensemble run with 20 realizations is done. 

Besides a deterministic coarse-grid simulation denoted as \emph{LR}, a deterministic high resolution (HR) simulation is performed that provides us with a reference solution. For all coarse model runs (both deterministic and stochastic), 
the resolution and parameters given in Table \ref{tab:params} are fixed to be the same. 
Note that Table \ref{tab:params} states the resolutions and parameters used for these various simulations.

\subsubsection*{Prediction of barotropic instability}

In this section, we compare the predictions of the barotropic instability for different coarse models to that provided by the HR reference simulation. The latter is obtained from the coarse-graining procedure through a bilinear interpolation of the high resolution snapshots. 

\begin{figure}
	\includegraphics[width=\textwidth]{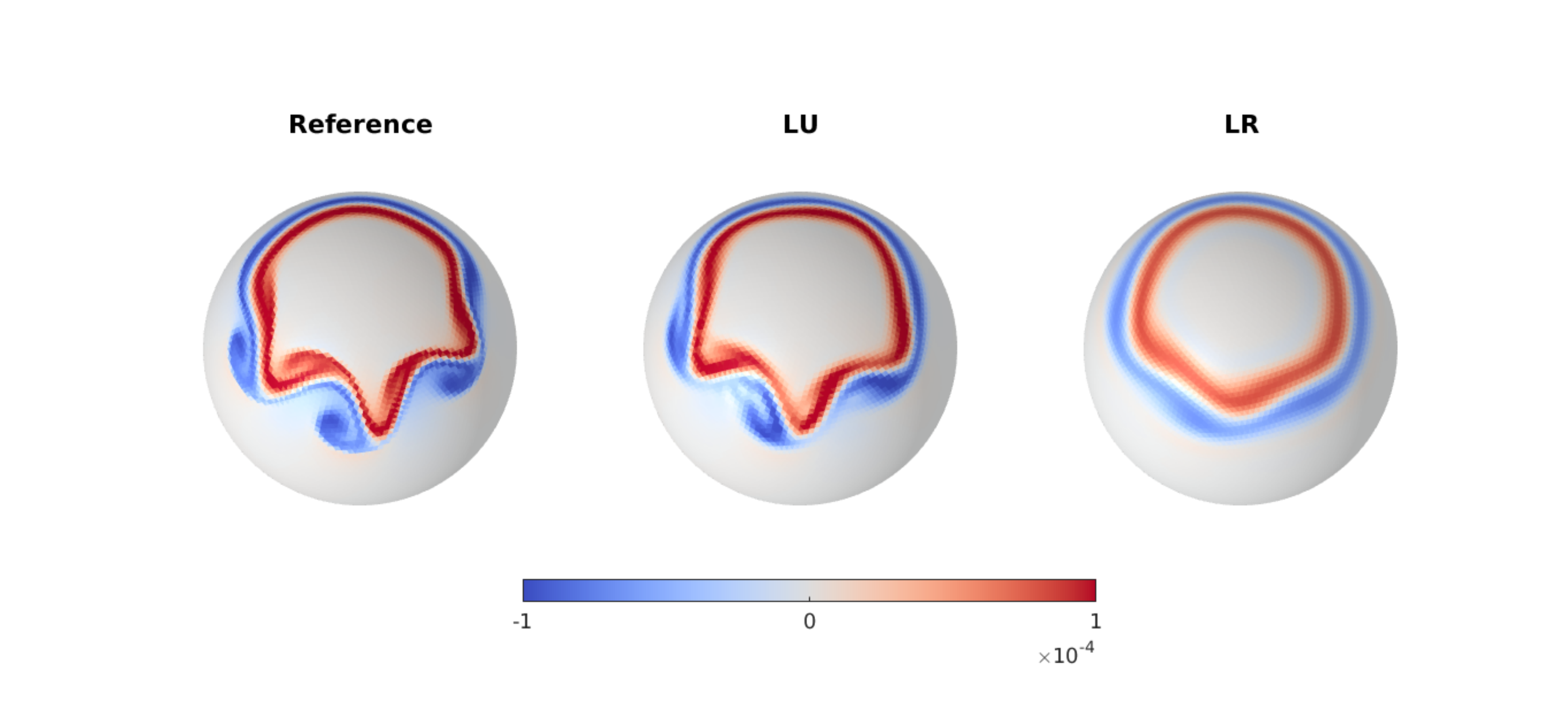}
	\caption{Snapshots of the vorticity field on the sphere for different models (with $20480$ triangles) after 5 days. From left to right: reference, ensemble mean of LU online and deterministic LR. For the simulations we use the parameters given in Table \ref{tab:params}.
	}
	\label{fig:snapOnSphere}
\end{figure}

\begin{figure}[htbp]
	\begin{center}
		\textbf{Day 4} \hspace{9.25em}
		\textbf{Day 5} \hspace{9.25em}
		\textbf{Day 6} \par\medskip
		\includegraphics[width=\textwidth]{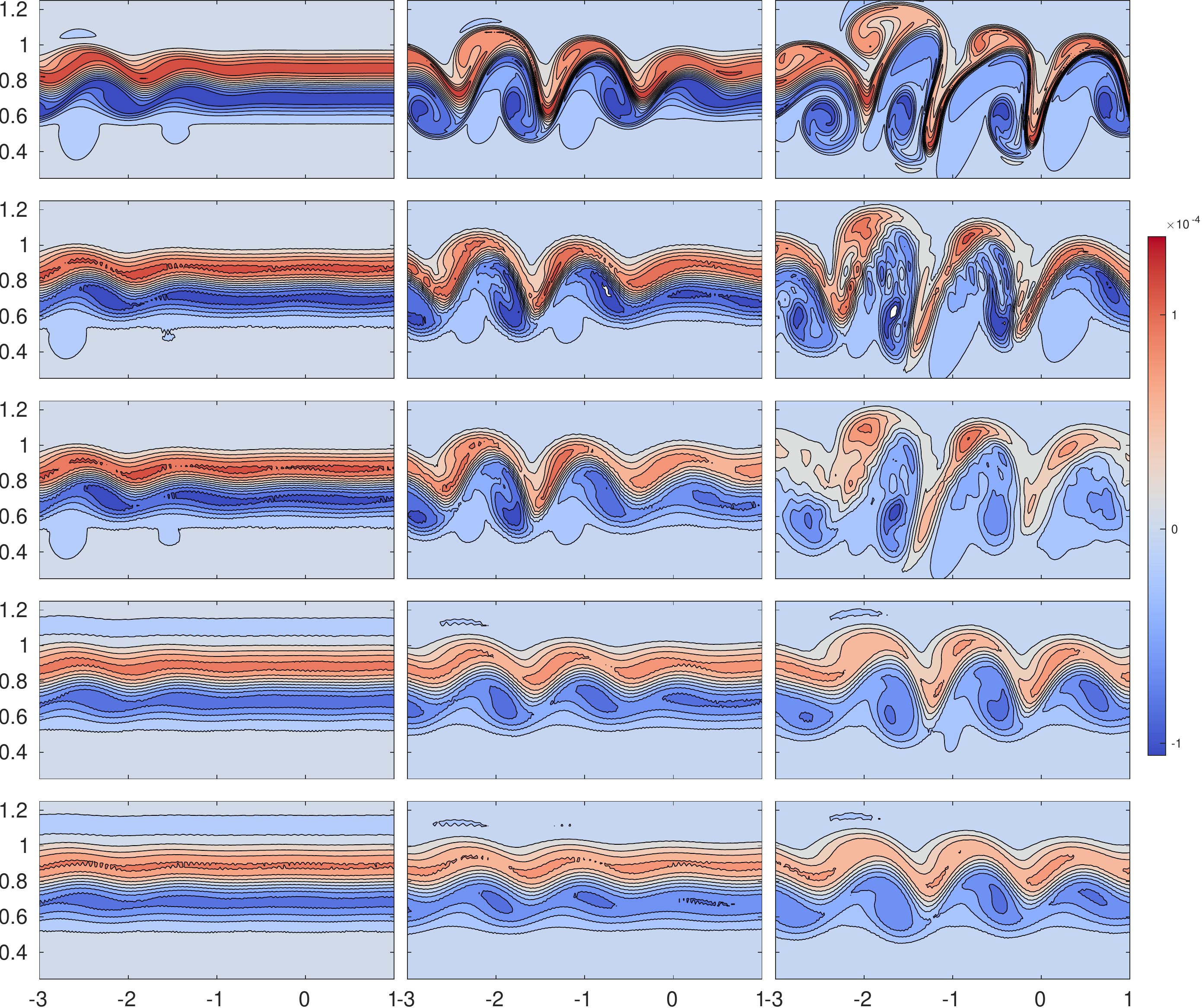}
		\caption{Comparison of the vorticity contour plots along the mid-latitude jet for different models (with $20480$ triangles) at day 4, 5 and 6 respectively. From top to bottom: reference, ensemble mean of LU on-line, ensemble mean of LU off-line, ensemble mean of PIC 1 and deterministic LR. The contour interval is fixed to $2 \times 10^{-5}\, \text{s}^{-1}$, the x-axis is longitude (in rad) and the y-axis is latitude (in rad). For the simulations we use the parameters given in Table \ref{tab:params}.
		}
		\label{fig:contour}
	\end{center}
\end{figure}

In Figure \ref{fig:snapOnSphere}, we illustrate  snapshots of the vorticity fields on the sphere for the reference, LU and deterministic models after a simulation time of 5 days. We can clearly see that the LU ensemble mean better captures the large-scale structure of the reference flow than the deterministic simulation. To better distinguish the differences in the simulations, contour plots of the vorticity fields at day 4, 5 and 6, localized at the mid-latitude of the sphere, are given in Figure \ref{fig:contour}. From the evolution of the reference vorticity fields we observe that the barotropic instability of the mid-latitude jet starts to develop at day 4. Subsequently, more and more small-scale features emerge and the flow becomes turbulent. Furthermore, both LU on-line and LU off-line simulations exhibit the stretched out wave at day 5 in the same way as the reference does, and that some big vortices start to separate from the wave at day 6. On the other hand, these characteristics are not correctly captured in both PIC 1 and LR simulations. We remark that the results of the PIC 2 simulations are not included in Figure \ref{fig:contour}, since they behave quite similarly to the PIC 1 runs.


To physically interpret the above results, it is useful to analyze the energy spectra of the different models. From a basic knowledge of the two-dimensional turbulence theory \cite{McWilliams2006GFD}, the potential enstrophy is transferred from the large scales to the small scales by the direct cascade, whereas the kinetic energy is transferred from the small scales to the large scales by the inverse cascade. However, introducing only a dissipation mechanism for coarse models often leads to an excessive decrease of the resolved kinetic energy \cite{Arbic2013, Kjellsson2017}. 

\begin{figure}[htbp]
	\begin{center}
		\hspace{3em}
		\textbf{Kinetic energy} \hspace{11em}
		\textbf{Normalized enstrophy} \par\medskip
		\includegraphics[width=\textwidth]{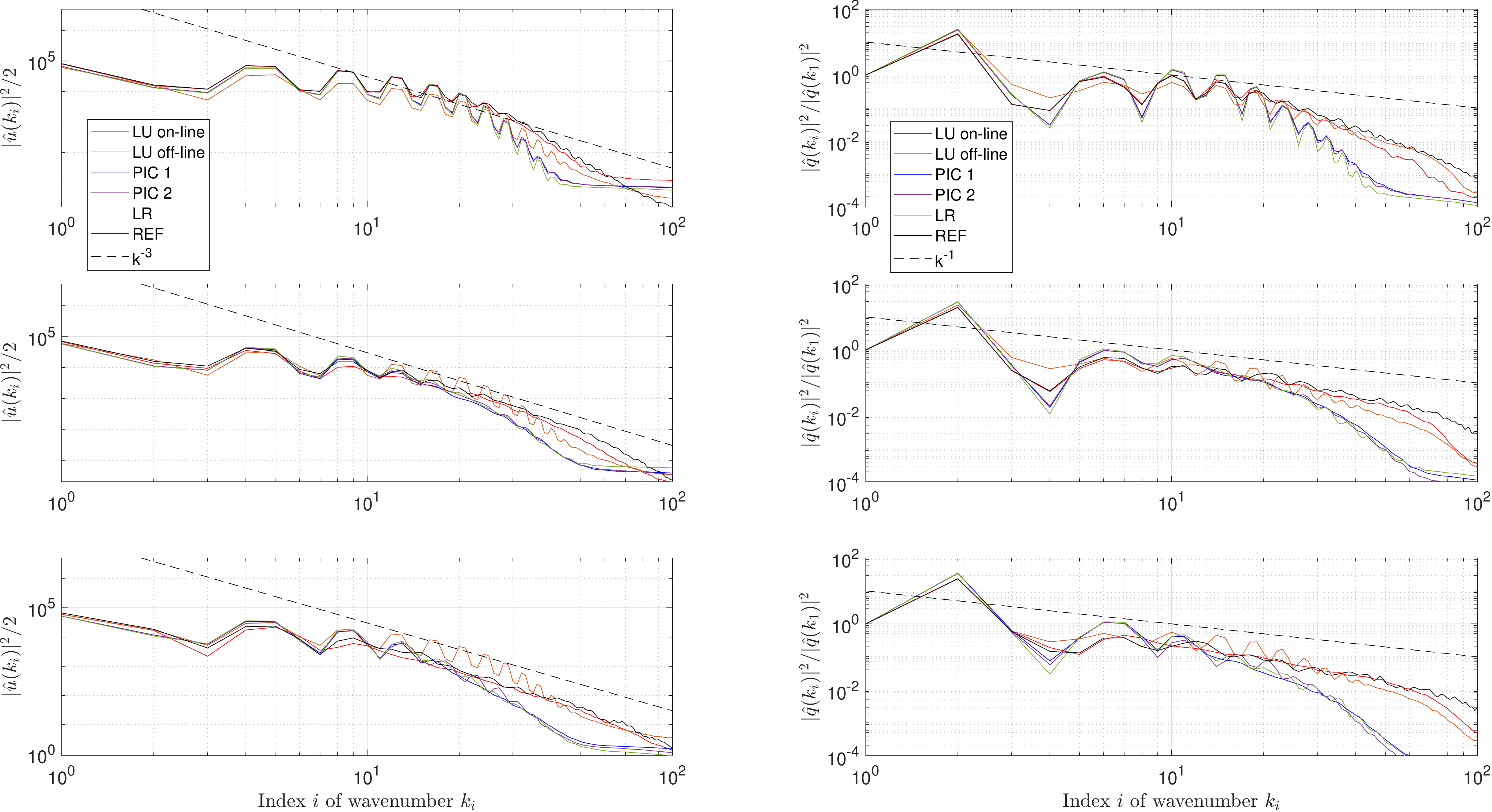}
	\end{center}
	\caption{Comparison of the ensemble mean of the kinetic energy (left column) spectrums and the potential enstrophy (right column) spectrums for different models (with $20480$ triangles) at day 5 (1st row), 7 (2nd row) and 10 (3rd row) respectively. Note that the potential enstrophy is defined by the square of the potential vorticity and 
		each potential enstrophy spectrum is normalized by its first value at the largest wavenumber. The dashed line is the $k^{-3}$ (left column) and $k^{-1}$ (right column) power law. These power laws for the RSW equations are discussed in \cite{ring02Ay,chen11Ay}.}
	\label{fig:specComp}
\end{figure}

In our test case, this kind of issue is present in both PIC and the LR simulations, where the small-scale energy and enstrophy are over-dissipated, as illustrated in Figure \ref{fig:specComp}. 
On the other hand, introducing the non-linear convection by the noise, the LU dynamical systems bring higher turbulent energy and enstrophy to the small scales, which leads to a better structuring of the large-scale flow. 
For instance, the time evolutions of the ensemble mean of the energy and enstrophy spectra for both LU on-line and LU off-line simulations are much closer to that of the references. 
However, the LU off-line spectrum changes little over time between wavenumbers 10 and 40 because the a priori obtained EOFs impose at each time step large scale modes on those scales. This is a drawback from a stationary noise.  
Note that these spectra on the sphere are calculated using the method proposed by \cite{Aechtner15}: first, the energy and enstrophy is interpolated onto a Gaussian grid, then the spherical harmonics basis are used to compute the power spectral density.

\subsubsection*{Evaluation of ensemble forecasts}

Once the ensembles have been produced by the random models, we measure the reliability of the ensemble forecast systems by some simple metrics. But before we do so, let us first demonstrate qualitatively the time evolution of each ensemble spread and compare it with the observation trajectory (obtained from the HR reference simulation). 
To determine the latter, we evaluate the local vorticity field of the reference at different grid points in the region of the mid-latitude jet. These points serve as observation points. 
The evolution of the spread of the ensemble forecast systems is then built by the $95\%$ confident interval of its ensemble trajectories at each selected point.

\begin{figure}[htbp]
	\begin{center}
		\includegraphics[width=\textwidth]{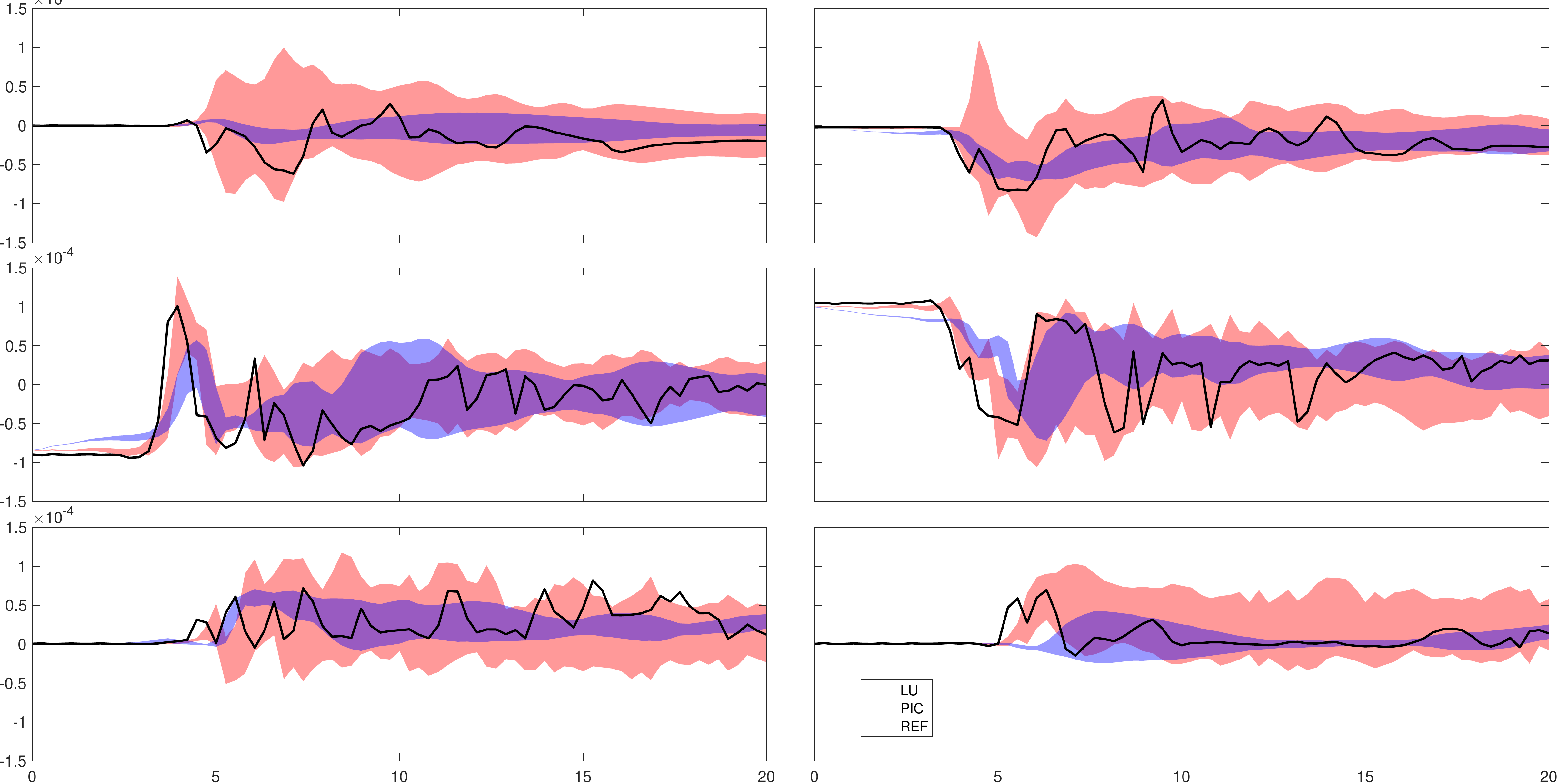}
	\end{center}
	\caption{Comparison of the ensemble spread evolution over 20 days of the vorticity field for the LU-offline (red area) runs and the PIC-offline (blue area) runs, at six different locations $\Theta=(0.4, 0.56, 0.72, 0.88, 1.04, 1.2)\, \text{rad}$ along the longitude $\Upsilon = - 1.53\, \text{rad}$. The observation trajectories are shown by the black lines. }
	\label{fig:spread}
\end{figure}

In Figure \ref{fig:spread} we compare the reference simulation and the simulations obtained from the off-line noise. To make the figure easier to read, only the  off-line noise is shown since the on-line noise behaves in a similar way. As shown, for the six local points chosen along the longitude $\Upsilon = -1.53\, \text{rad}$, the ensemble spreads of the LU off-line system are large enough to almost always include the observation trajectories, whereas the spreads of the PIC 1 system are quite small so that the observations are not always contained within the spread. 
For the latter, this will result in a wrong coupling of the measurement and the ensemble system, when performing data assimilation \cite{Gottwald2013, Franzke2015}.

To quantify whether the ensemble spread of the forecast system represents the true uncertainty of the observations (obtained from the reference simulation), the rank histogram \cite{Talagrand1997, Hamill2001} is widely adopted as a diagnostic tool. 
This approach checks where the verifying observation usually falls w.r.t. the ensemble forecast states which are arranged in an increasing order at each grid point. In an ensemble with perfect spread, each member represents an equally likely scenario, so the observation is equally likely to fall between any two members. To construct the rank histogram in our test case, we proceed as follows:
\begin{enumerate}
	\item At every grid point $\vec{x}_i$, we rank the $N_e$ vorticity values $\{ q^{(j)} (\vec{x}_i) \}_{\sub j=1,\ldots,N_e}$ of the ensemble from lowest to highest. This results in $N_e+1$ possible bins which the observations can fall into, including the two extremes;
	\item Identify which bin the observation vorticity $q^o (\vec{x}_i)$ falls into at each point $\vec{x}_i$;
	\item Tally over all observations $\{ q^o (\vec{x}_i) \}_{\sub i=1,\ldots,N_o}$ to create a histogram of rank.
\end{enumerate}

\begin{figure}[htbp]
	\begin{center}
		\hspace{3.25em}
		\textbf{Day 5} \hspace{5.5em}
		\textbf{Day 10} \hspace{5.5em}
		\textbf{Day 15} \hspace{5.5em}
		\textbf{Day 20} \par\medskip
		\includegraphics[width=\textwidth]{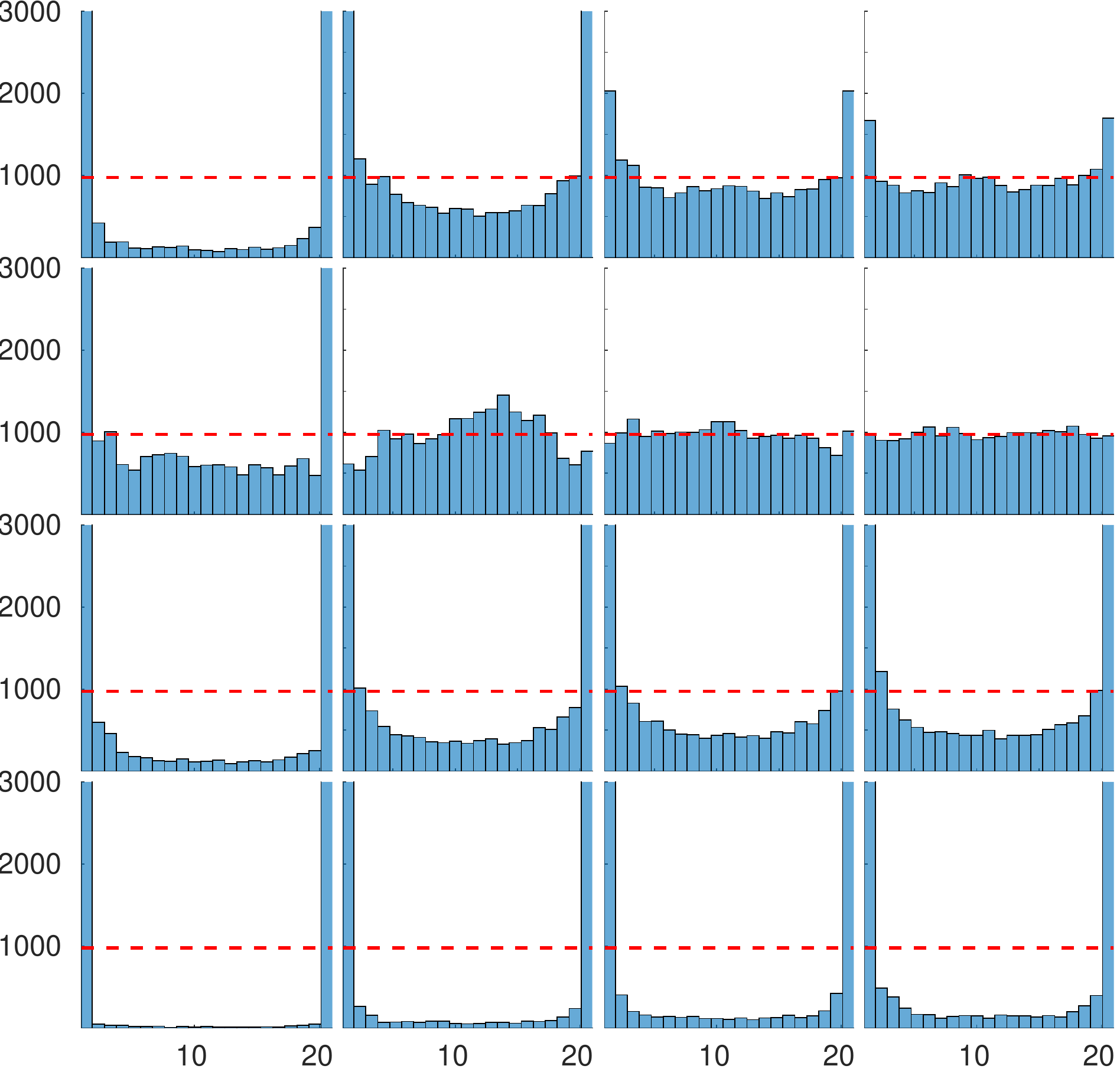}
	\end{center}
	\caption{Comparison of the rank histograms for the LU on-line (1st row) runs, the LU off-line (2nd row) runs, the PIC 2 (3rd row) runs and PIC 1 (last row) runs, at day 5, 10, 15 and 20 respectively.}
	\label{fig:hist}
\end{figure}

As shown in Figure~\ref{fig:hist}, the histograms of both random models exhibit a U-shape for a few days in the beginning, 
while after a simulation time of about 10 days, the histograms of both LU on-line and LU off-line systems become mostly flat. A U-shape indicates that the ensemble spread is too small so that many observations are falling outside of the extremes of the ensemble while a dome-shape indicates the contrary. A flat histogram, in contrast, indicates that the ensemble members and observations are sampled from a common distribution. We observe that the LU off-line system performs slightly better than the LU on-line version. In contrast to these very good ensemble spreads, the histograms of both PIC 2 and PIC 1 systems remain in a U-shape during the entire simulation period which indicates that these systems do not accurately estimate the correct uncertainty around the observations.

It is important to notice that a flat rank histogram does not necessarily imply good forecasts, it only measures whether the observed probability distribution is well represented by the ensemble. To verify that a forecast is reliable, we need more criteria. 
One necessary criterion \cite{Weigel2012} for a reliable ensemble forecast is that the mean squared error (MSE) of the ensemble matches the mean intra-ensemble variance (MEV), up to an ensemble size-dependent scaling factor, {\em i.e.} 

	\begin{align}\label{eq:msbmev}
	\text{MSE}\, (t) &= \frac{1}{N_o} \sum_{i=1}^{N_o} \big( q^{o} - \widehat{\Exp} [q] \big)^2  (t, \vec{x}_i) \nonumber\\
	&\approx \Big( \frac{N_e+1}{N_e} \Big) \frac{1}{N_o} \sum_{i=1}^{N_o} \widehat{\text{Var}} [q]  (t, \vec{x}_i) = \frac{N_e+1}{N_e}\, \text{MEV}\, (t),
	\end{align}

where $\widehat{\Exp} [q] = \frac{1}{N_e} \sum_{j=1}^{N_e} q^{(j)}$ and $\widehat{\text{Var}} [q] = \frac{1}{N_e-1} \sum_{j=1}^{N_e} \big( q^{(j)} - \widehat{\Exp} [q] \big)^2$ denote the empirical mean and the empirical variance, respectively.

\begin{figure}[htbp]
	\begin{center}
		\includegraphics[width=0.55\textwidth]{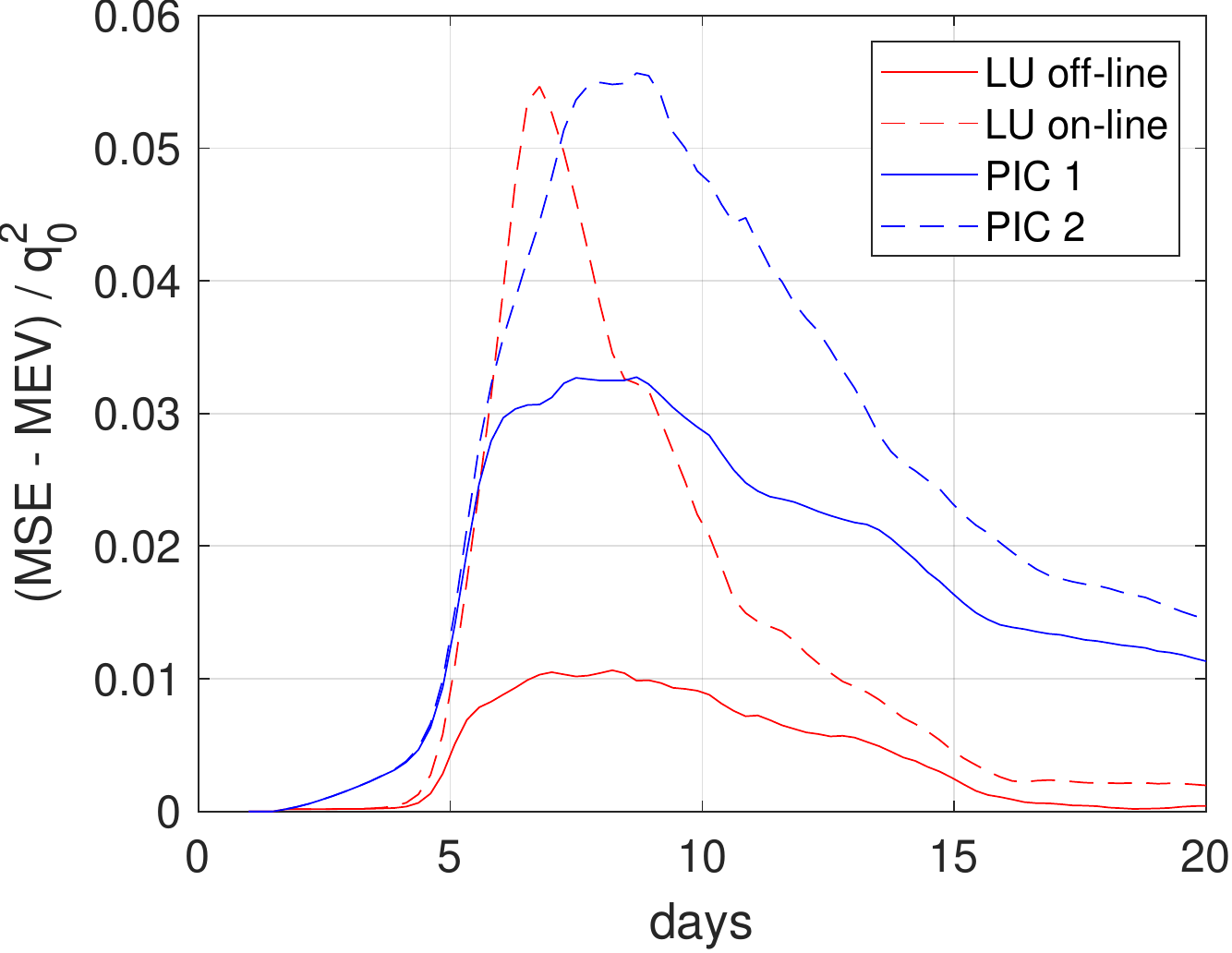}
	\end{center}	
	\caption{Comparison of the differences between the mean square error (MSE) and the mean ensemble variance (MEV) of the ensemble vorticity fields for the LU on-line (red dashed line) runs, the LU off-line (red solid line) runs, the PIC 2 (blue dashed line) runs and the PIC 1 (blue solid line) runs. Note that these differences are normalized by $q_0=\| q (\Upsilon, \Theta, t=0) \|_{\infty}$. 
	}
	\label{fig:MSE}
\end{figure}

In Figure \ref{fig:MSE}, we compare the differences in time between the MSE and the MEV, normalized by the squared maximum of the initial vorticity, for the different random models from above. 
From these curves we can deduce that the LU off-line system exhibits the lowest errors during the entire simulation time of 20 days. In particular, during the first 10 days, these errors are significantly lower when compared to the other models, which can be explained by the fact that the LU off-line system incorporates data from the reference into the ensemble, which increases the reliability of the ensemble forecast. Although the errors between MSE and MEV of the LU on-line system is larger than the LU offline
system from day 5 to day 10, they remain at low level from day 10 onwards, implying that 
the reliability of the former increases for longer simulation times. 
In contrast, both PIC 1 and PIC 2 systems show higher error values at most of the times and hence provide less reliable ensembles. We remark that other metrics, such as the continuous ranked probability score \cite{Resseguier2020arcme, Weigel2012}, can also be used to measure a calibrated ensemble.


\section{Conclusions}\label{sec:conclusions}
In this study, we introduced a stochastic version of the rotating shallow water equations under location uncertainty (RSW-LU). The derivation is based on a stochastic Reynolds transport theorem, where the fluid flow is decomposed into a large-scale component and a noise term modelling the unresolved small-scale flow. A benefit of this approach is that the total energy is conserved along time for any realization. 
In order to preserve this structure, we combined an energy (in space) preserving discretization of the underlying deterministic equations of this RSW--LU system with approximations of the stochastic terms that are based on standard finite volume/difference operators.

We could show for an f-plane test case that this approach leads for homogeneous noise to a discretization of the RSW--LU system that preserves (spatially) the total energy. Moreover, using inhomogeneous noise that well captures the impact of small scales on the large-scale flow, we demonstrated that for a barotropically unstable jet on the sphere our proposed RSW--LU model better predicts the development of the instabilities than a comparable deterministic model, while the ensemble spread of the RSW--LU system is more likely to contain the observations compared to an ensemble of deterministic simulations with perturbed initial conditions (PIC). We also showed that the RSW--LU forecast systems follows a common distribution of the observations and is more reliable than the PIC system.

Showing accurate ensemble spreads and reliable forecasting skills, we will next apply our developed RSW--LU system to data assimilation. We will also work towards discretizations of stochastic flow models in the framework of LU that preserve total energy both in space and time to which the present work provides a first step.
Exploiting the modular approach of combining different discretizations for deterministic and stochastic terms, in future work we will explore the possibility to consistently extend existing atmospheric and ocean models with stochastic parametrizations. 
We remark that the stochastic approach proposed in this work could be extended to arbitrary Riemannian manifold. In this setting, it would be easier to first convert the It\^{o} integrals to the Stratonovich representations (see Remark 2), and then transform the latter from Euclidean space to other subspaces of Riemannian manifold under diffeomorphism \cite{Hsu2002}. This application could be helpful for the deep atmosphere component of various global numerical weather prediction and climate models, where the domain significantly differs from Euclidean space.

\section*{Acknowledgments}
The authors acknowledge the support of the Mitacs Globalink Research Award and of the ERC EU project 856408-STUOD. The author Werner Bauer would like to acknowledge funding from NERC NE/R008795/1. Besides, we would like to thank Alexander Bihlo and Scott MacLachlan for helpful discussions and thank Matthias Achtner for providing code to compute the energy spectrum on the sphere. We also gratefully acknowledge the reviewers for their insightful comments and suggestions that helped us to significanlty improve this manuscript.
\\
The code to reproduce the results  is available at  \url{https://github.com/RudigerBrecht/RSW-LU}
\cite{rudigerbrecht_2021_4884919}.
The scripts and data to reproduce the figures can be obtained from \url{https://zenodo.org/record/5576233}.

\appendix

\section{Energy conservation of RSW--LU}\label{A:energy}


This appendix demonstrates the energy conservation of the RSW--LU system \eqref{seq:RSWLU}. 
Let us recall that the density of the kinetic energy (KE) and of the potential energy (PE) of the large-scale flow in the shallow water system \cite{Vallis2017} is, respectively, given by

	\begin{subequations}\label{seq:E-RSW}
		\begin{align}
		\KE &= \int_{0}^{h} \frac{\rho_0}{2} |\vec{u}|^2\, \df z = \frac{\rho_0}{2} h |\vec{u}|^2, \label{eq:KE} \\
		\PE &= \int_{0}^{h} \rho_0 g z\, \df z = \frac{\rho_0}{2} g h^2, \label{eq:PE}
		\end{align}
	\end{subequations}

where $|\vec{u}|^2 = \vec{u} \bdot \vec{u}$ and we assume that $\rho_0 = 1$ and the bottom is flat, {\em i.e.} $\eta_b = 0$ for algebraic simplicity. 
In order to explain the conservation of energy more concisely, we adopt the following product rule of the stochastic transport operator as derived in \cite{Resseguier2017gafd1}. For scalar tracers $f, g$ transported by the stochastic flow and incorporating smooth-in-time external forcings $F, G$, i.e. $\Df_t f = F\, \dt$ and $\Df_t g = G\, \dt$, we have

	\begin{equation}
	\Df_t (f g) = g \Df_t f + f \Df_t g.
	\end{equation}

Applying this rule to the definition of PE \eqref{eq:PE} and using the mass equation \eqref{eq:RSWLU-mass}, the PE evolution reads
\begin{subequations}
	
		\begin{equation}
		\Df_t \PE = g h \Df_t h = - g h^2 \div \vec{u}\, \dt = - 2\PE \div \vec{u}\, \dt.
		\end{equation}
	
	Similarly, from both mass equation and momentum equation in \eqref{seq:RSWLU}, noting that $\vec{u} \bdot (\vec{f} \times \vec{u}) = 0$ and recalling that $\eta_b =0$, we derive the evolution of KE \eqref{eq:KE}:
	
		\begin{align}
		\Df_t \KE &= h \vec{u} \bdot \Df_t \vec{u} + \alf |\vec{u}|^2 \Df_t h \nonumber\\ 
		&= - \alf \vec{u} \adv \big( g h^2 \big)\, \dt - \alf h |\vec{u}|^2 \div \vec{u}\, \dt
		= -\big( \vec{u} \adv \PE +  \KE \div \vec{u} \big)\, \dt .		
		\end{align}
	
\end{subequations}
Subsequently, we deduce the evolution  of the total energy density $\E = \KE + \PE$,
\begin{subequations}
	
		\begin{equation}
		\Df_t \E  = - \big( \div (\vec{u}\, \PE) + \E \div \vec{u} \big)\, \dt .
		\end{equation}
	
	Expanding the stochastic transport operator \eqref{eq:STO}, and including the incompressible constraints \eqref{eq:RSWLU-incomp}, the previous equation can be re-written as
	
		\begin{equation}\label{eq:dt-Eden}
		\df_t \E = - \div \Big( \big( \E\, (\vec{u} - \vec{u}_s) + \PE\, \vec{u} - \alf \vec{a} \grad \E \big)\, \dt + \E\, \sdbt \Big).
		\end{equation}
	
\end{subequations}
Let us now assume some ideal boundary conditions for the resolved and unresolved components:

	\begin{equation}\label{eq:cond-bord}
	\vec{u}\bdot \vec{n}\, \big|_{\partial \Omega} = \vec{u}_s \bdot \vec{n}\, \big|_{\partial \Omega} = \sdbt \bdot \vec{n}\, \big|_{\partial \Omega} = 0,
	\end{equation}

where $\partial \Omega$ denotes the boundary of the fluid domain $\Omega$ and $\vec{n}$ stands for the outward pointing unit normal. Combining Equations \eqref{eq:dt-Eden} and \eqref{eq:cond-bord}, one can show that the total energy (integration of energy density over domain) is invariant over time:

	\begin{equation}
	\df_t \int_{\Omega} \E (\vec{x},t) \df\vec{x} = - \int_{\partial \Omega} \Big( \big( \E\, (\vec{u} - \vec{u}_s) + \PE\, \vec{u} - \alf \vec{a} \grad \E \big)\, \dt + \E\, \sdbt \Big) \bdot \vec{n}\, \df l  = 0,
	\end{equation}

in which the following argument is used

	\begin{equation}
	\vec{n} \bdot (\vec{a} \grad \E)\, \dt = \sum_{i,j=1,2} n_i (a_{i,j}\, \dt) \partial_{x_j} \E = \sum_{j=1,2} \Exp \big[ \underbrace{\sum_{i=1,2} n_i (\sigma \df B_t)^i}_{=\, 0\ \text{at}\ \partial \Omega}  (\sigma \df B_t)^j \big] \partial_{x_j} \E.
	\end{equation}


\section{Parameterizations of noise}\label{sec:param-noise}


This section describes briefly some existing parametrization methods for the noise structure. For interested readers, more technical details can be found in \cite{Resseguier2020arcme}.


\subsection{Homogeneous noise}\label{sec:fft-noise}
From Definitions \eqref{eq:corr} and \eqref{eq:var0}, a homogeneous noise means that its correlation operator $\vec{\sigma}$ is a convolution operator and the variance tensor $\vec{a}$ reduces to a constant matrix. To ensure the incompressible constraint \eqref{eq:RSWLU-incomp} of a two-dimensional noise, \cite{Resseguier2017gafd2} proposed an isotropic model defined through a random stream function

	\begin{equation}\label{eq:isonoise}
	\vec{\sigma} (\vec{x})\, \df \vec{B}_t = \gradp \big( \breve{\varphi} \star \df B_t \big) (\vec{x}),
	\end{equation}

where $\gradp = [-\partial_y, \partial_x]\tp$ denotes the perpendicular gradient and $\breve{\varphi} \star \df B_t$ stands for the random stream function with a convolution kernel $\breve{\varphi}$ (and the symbol $\star$ denotes a convolution). Both isotropy and incompressibility of the noise result in a (constant) diagonal variance tensor $a_{\sub 0} \Id_{\sub 2}$ with the eddy-viscosity-like coefficient $a_{\sub 0}$ and the two-dimensional identity matrix $\Id_{\sub 2}$. 
For the current work, the divergence-free constraint of the ISD in Equation \eqref{eq:RSWLU-incomp} is thus naturally satisfied. 
In practice, the convolution kernel $\breve{\varphi}$ is specified by three parameters: a fixed omni-directional spectrum slope $s$, a band-pass filter $f_{\sub\text{BP}}$ with support in the range of two wavenumbers $\kappa_{\sub m}$ and $\kappa_{\sub M}$, and the coefficient $a_{\sub 0}$. In fact, the Fourier transform of the random stream function $\breve{\varphi} \star \df B_t$ can be defined as:

	\begin{equation}
	\widehat{\breve{\varphi} \star \df B_t} (\vec{k}) \defin \frac{A}{\sqrt{\Delta t}} f_{\sub\text{BP}} \left(\left\| \vec{k} \right\|\right) \left\| \vec{k} \right\|^{-\alpha} \widehat{\xi_t} (\vec{k})\ \text{with}\ \alpha = (3+s)/2,
	\end{equation}

where $\widehat{\bullet}$ denotes the Fourier transform coefficient, $\xi_t$ is a space-time white noise, and $A$ is a constant to ensure $\Exp \bigl\| \sdbt \bigr\|^2 = 2 a_{\sub 0} \Delta t$ (see Equations \eqref{eq:cov} and \eqref{eq:var0}) with $\Delta t$ the size of one time stepping and $\Exp$ the expectation operator. In the simulations, the maximal wavenumber $k_{\sub M}$ of the noise can usually be chosen as the effective resolution cutoff, the minimal wavenumber can be set to $k_{\sub m} = k_{\sub M}/2$, and the theoretical spectrum slope of a two-dimensional flow is given by $s = -3$. 
{Note that on the sphere homogeneous noise could be generated through spherical harmonics.}


\subsection{Heterogeneous noise}\label{sec:eof-noise}

In the following, two parameterizations of the heterogeneous noise are presented. These approaches result from the spectral decomposition \eqref{seq:KL} used to construct the EOFs of the covariance. However in practice, we work with the Eulerian velocity rather than with the Lagrangian displacement. 

\subsubsection{Off-line learning of EOFs}

The first method consists in calibrating EOFs from the off-line simulation data with the EOFs assumed to be time-independent. To this end, let us consider a set of velocity snapshots $\{ \vec{u}_{\text{o}} (\vec{x},t_i) \}_{i=1,\ldots,N_t}$, that have been {\em a priori} coarse-grained from high-dimensional data. 
Applying 
the singular value decomposition (SVD) for the fluctuations $\vec{u}_{\text{o}}' = \vec{u}_{\text{o}} - \overline{\vec{u}}_{\text{o}}$ (where  $\overline{\bullet}$ denotes a temporal average) enables us to build a set of EOFs $\{ \vec{\phi}_i \}_{i=1,\ldots,N_t}$. In addition, we suppose that the fluctuations of the large-scale flow live in a subspace spanned by $\{ \vec{\phi}_i \}_{i=1,\ldots,m-1}$ (with $m < N_t$) and that the small-scale random drift $\sdbt / \Delta t$ lives in the complemented subspace spanned by  $\{ \vec{\phi}_i \}_{i =m, \ldots,  N_t}$ such that

	\begin{equation}\label{eq:offline-noise}
	\frac{1}{\Delta t} \vec{\sigma} (\vec{x})\, \df \vec{B}_t = \sum_{i = m}^{N_t} \sqrt{\lambda_i} \vec{\phi}_i (\vec{x}) \xi_i,\ \quad 
	\frac{1}{\Delta t} \vec{a} (\vec{x}) = \sum_{i = m}^{N_t} \lambda_i \vec{\phi}_i (\vec{x}) \vec{\phi}_i\tp (\vec{x}),
	\end{equation}

where $\lambda_i$ is the eigenvalue associated to the spatial mode $\vec{\phi}_i$ and $\xi_i$ is a standard Gaussian variable. In practice, there exists an open question in \eqref{eq:offline-noise}, that is how to adequately choose the ``splitting mode'' $\vec{\phi}_m$. Recently, \cite{Bauer2020ocemod} proposed to fix it by comparing the time-averaged energy spectrum of the observations and the one from a coarse-grid deterministic simulation.

\subsubsection{On-line learning of EOFs}\label{sec:online}
The previously described data-driven calibriation of EOFs is a quite efficient procedure. However, such observation data 
are not always available. \cite{Bauer2020jpo, Resseguier2020arcme} proposed an alternative approach in which some local fluctuations, called \emph{pseudo-observations} (PSO), are generated directly from a coarse-grid simulation. Then, the SVD is applied on those PSO to estimate a set of EOFs such that the noise associated with its variance tensor will be built in the same way as in \eqref{eq:offline-noise}. Finally, the magnitude of the noise and variance should be scaled down to smaller scales based on a similarity analysis. 

The approach proposed first defines $N_o$ PSO (denoted as $\vec{u}'$) at each grid point. For a given time $t$ and a current coarse velocity $\vec{u}$, we build the PSO by sliding a local window of size $N_w \times N_w$ over the spatial grid (with $N_w$ the grid number in one direction of the local window). We denote the spatial scale of the window by $L = N_w l$, where $l$ is the smallest scale of the simulation. At every grid point $\vec{x}_{i,j}$, we list the $N_w^2$ velocity values contained in the window centered at that point:

	\begin{equation}
	I(\vec{x}_{i,j},t) \defin
	\left \{ \vec{u} ( \vec{x}_{p,q} ,t ) \bigg | |p - i| \leq \frac{N_w - 1}{2} , |q - j| \leq \frac{N_w  - 1}{2} \right \}.
	\end{equation}

Note that appropriate boundary conditions (replication, periodicity, etc.) are adopted when looking at a point on the border. Then, independently for each $n \in \{1, \ldots, N_o\}$ and for each point $\vec{x}_{i,j}$, we set the value of the PSO $\vec{u}' (\vec{x}_{i,j},t,n)$ by randomly choosing a value in the set $ I(\vec{x}_{i,j},t)$. After this, we average over the realization index $n$ to build an empirical covariance. Then, from the SVD we obtain a set of EOFs $\{ \vec{\phi}^{\sub (L)}_i \}_{i = 1, \ldots, N_o}$, and a spectral representation of the small-scale velocity:
\begin{subequations}
	
		\begin{equation}\label{eq:realization-2}
		\frac{1}{\Delta t} \vec{\sigma}^{\sub (L)} (\vec{x},t)\, \df{\vec{B}}_t = \sum_{i=1}^{N_o} \vec{\phi}_i^{\sub (L)} (\vec{x},t) \xi_i .
		\end{equation}
	
	Since the PSO $\vec{u}'$ have been generated at a spatial scale of the window $L = N_w l$, they must be scaled down to the ``simulation scale'' $l$.  In 3D, according to an auto-similarity assumption of the velocity fluctuations  \cite{Kadri2017}, the small-scale flow $\vec{\sigma}^{\sub (l)}  \df{\vec{B}}_t$ associated with its variance tensor $\vec{a}^{\sub (l)}$ can be rescaled as
	
		\begin{equation}\label{eq:al}
		\vec{\sigma}^{\sub (l)}  \df{\vec{B}}_t = \left(\frac{l}{L}\right)^{1/3} \ \vec{\sigma}^{\sub (L)} \df{\vec{B}}_t,\ \quad \vec{a}^{\sub (l)} = \left(\frac{l}{L}\right)^{2/3} \vec{a}^{\sub (L)}.
		\end{equation}
	
\end{subequations}
In our case, noting that the small-scale fluctuations are still 3D (even though the vertical component is not known), we keep the same scaling.  As shown in Section \ref{sec:Galeswky}, such flow-dependent noise has a good performance in long-term simulation, yet the drawback is that the computational costs are significantly higher compared to the previous off-line procedure, as the SVD is computed at each time step.

\section{Discretization of LU terms}\label{A:LUdiscrete}
%

Starting with a given predicted velocity vector with edge values $V_{ij}$, we first have to reconstruct 
the full velocity vector field from these normal values. We use the reconstruction of the vector field 
in the interior of each triangle proposed by \cite{perot2006mimetic}:

	\begin{equation}\label{eq:recvelocity}
	\vec{u}_i= \frac{1}{|T_i|}\sum_{k=j, i_-, i_+}|e_{ik}| (\vec{x}^{e_{ik}}-\vec{x}^{T_i}) V_{ik},
	\end{equation}

where $\vec{x}^{e_{ik}}$ are the coordinates of the edge midpoint and $\vec{x}^{T_i}$ are the coordinates of the triangle circumcentre. 
By averaging values from neighboring triangles, we obtain  the corresponding values at the edge midpoints 
or vertices (see \cite{bauer2013toward} for details).

This reconstructed velocity vector field will be used to generate the noise as described in \ref{sec:param-noise}.
After the noise has been constructed on the Cartesian mesh, we evaluate the 
discrete noise vector ${(\sdbt)}_{ij}$ and the discrete variance tensor $(\vec{a})_{ij}$ at the triangle edge midpoints. 
This information will then be used to calculate the LU noise terms 
in \eqref{eq:discreteMomentumLU} and \eqref{eq:discreteMassLU}.

To calculate the derivatives in these stochastic terms, we use the normal and tangential gradient operators,
i.e. the gradient operator of \eqref{eq:discreteGrad}. To use it, we have to average values, e.g. the term $(a_{kl}F)$, 
to cell centers and vertices and the resulting differential will be an expression located at the edge midpoint. 
In more detail, we can represent the partial derivative in Cartesian coordinates by 

	\begin{equation}\label{eq:discretePartial}
	(\partial_{x_l}F)_{ij}= (\text{Grad}_n~F)n^l_{ij}+ (\text{Grad}_t~F)t^l_{ij}, \qquad \qquad l=1, 2.
	\end{equation}

Concretely, to discretize \eqref{eq:rewriteDiffu},  we first compute $(\partial_{x_l}(a_{kl}F))_{ij}$ using Equation \eqref{eq:discretePartial}.
The subindex $ij$ indicates that the resulting term is associated 
to the edge midpoint. To apply the second derivative in \eqref{eq:rewriteDiffu}, i.e. 
$\left(\partial_{x_k} \left( \partial_{x_l} (a_{kl}F)\right)_{ij} \right)_{ij}$, we
proceed analogously, i.e. we first average the terms describing the first derivative to 
cells and vertices and then apply once more Equation \eqref{eq:discretePartial}.
We proceed similarly to represent the term $\nabla F$ in \eqref{eq:secondtermLU}.

\bibliography{mybib}

\begin{thebibliography}{79}
\providecommand{\natexlab}[1]{#1}
\providecommand{\url}[1]{\texttt{#1}}
\expandafter\ifx\csname urlstyle\endcsname\relax
  \providecommand{\doi}[1]{doi: #1}\else
  \providecommand{\doi}{doi: \begingroup \urlstyle{rm}\Url}\fi

\bibitem[Aechtner et~al.(2015)Aechtner, Kevlahan, and Dubos]{Aechtner15}
M.~Aechtner, N.~K.-R. Kevlahan, and T.~Dubos.
\newblock A conservative adaptive wavelet method for the shallow-water
  equations on the sphere.
\newblock \emph{Quarterly Journal of the Royal Meteorological Society},
  141\penalty0 (690):\penalty0 1712--1726, 2015.
\newblock \doi{10.1002/qj.2473}.

\bibitem[Anderson and Anderson(1999)]{Anderson1999}
J.~Anderson and S.~Anderson.
\newblock A {M}onte {C}arlo implementation of the nonlinear filtering problem
  to produce ensemble assimilations and forecasts.
\newblock \emph{Monthly Weather Review}, 127\penalty0 (12):\penalty0
  2741--2758, 1999.

\bibitem[Andrews and McIntyre(1978)]{Andrews78}
D.~Andrews and M.~McIntyre.
\newblock An exact theory of nonlinear waves on a {L}agrangian-mean flow.
\newblock \emph{Journal of Fluid Mechanics}, 89\penalty0 (4):\penalty0
  609--646, 1978.

\bibitem[Arbic et~al.(2013)Arbic, Polzin, Scott, Richman, and
  Shriver]{Arbic2013}
B.~K. Arbic, K.~L. Polzin, R.~B. Scott, J.~G. Richman, and J.~F. Shriver.
\newblock On eddy viscosity, energy cascades, and the horizonal resolution of
  gridded stallite altimeter products.
\newblock \emph{Journal of Physical Oceanography}, 43\penalty0 (2):\penalty0
  283--300, 2013.

\bibitem[Bauer(2013)]{bauer2013toward}
W.~Bauer.
\newblock \emph{Toward goal-oriented {R}-adaptive models in geophysical fluid
  dynamics using a generalized discretization approach}.
\newblock PhD thesis, Hamburg University Hamburg, 2013.

\bibitem[Bauer and Gay-Balmaz(2019{\natexlab{a}})]{Bauer2019}
W.~Bauer and F.~Gay-Balmaz.
\newblock Towards a geometric variational discretization of compressible
  fluids: the rotating shallow water equations.
\newblock \emph{Journal of Computational Dynamics}, 6:\penalty0 1,
  2019{\natexlab{a}}.

\bibitem[Bauer and Gay-Balmaz(2019{\natexlab{b}})]{Bauer2019a}
W.~Bauer and F.~Gay-Balmaz.
\newblock Variational integrators for anelastic and pseudo-incompressible
  flows.
\newblock \emph{Journal of Geometric Mechanics}, 11\penalty0 (4):\penalty0
  511--537, 2019{\natexlab{b}}.

\bibitem[Bauer et~al.(2020{\natexlab{a}})Bauer, Chandramouli, Chapron, Li, and
  M\'{e}min]{Bauer2020jpo}
W.~Bauer, P.~Chandramouli, B.~Chapron, L.~Li, and E.~M\'{e}min.
\newblock Deciphering the role of small-scale inhomogeneity on geophysical flow
  structuration: a stochastic approach.
\newblock \emph{Journal of Physical Oceanography}, 50\penalty0 (4):\penalty0
  983--1003, 2020{\natexlab{a}}.

\bibitem[Bauer et~al.(2020{\natexlab{b}})Bauer, Chandramouli, Li, and
  M\'{e}min]{Bauer2020ocemod}
W.~Bauer, P.~Chandramouli, L.~Li, and E.~M\'{e}min.
\newblock Stochastic representation of mesoscale eddy effects in
  coarse-resolution barotropic models.
\newblock \emph{Ocean Modelling}, 151:\penalty0 101646, 2020{\natexlab{b}}.

\bibitem[Berge et~al.(1987)Berge, Pomeau, and Vidal]{Berge-Pomeau-Vidal}
P.~Berge, Y.~Pomeau, and C.~Vidal.
\newblock \emph{Order within Chaos: Towards a Deterministic Approach to
  Turbulence.}
\newblock John Wiley \& Sons, New York, 1987.

\bibitem[Berner and Coauthors(2017)]{Berner-17}
J.~Berner and Coauthors.
\newblock Stochastic parameterization: Toward a new view of weather and climate
  models.
\newblock \emph{Bulletin of the American Meteorological Society}, 98:\penalty0
  565--588, 2017.

\bibitem[Bonaventura and Ringler(2005)]{bonaventura2005analysis}
L.~Bonaventura and T.~Ringler.
\newblock Analysis of discrete shallow-water models on geodesic delaunay grids
  with {C}-type staggering.
\newblock \emph{Monthly Weather Review}, 133\penalty0 (8):\penalty0 2351--2373,
  2005.

\bibitem[Brecht et~al.(2019)Brecht, Bauer, Bihlo, Gay-Balmaz, and
  MacLachlan]{Brecht2019}
R.~Brecht, W.~Bauer, A.~Bihlo, F.~Gay-Balmaz, and S.~MacLachlan.
\newblock Variational integrator for the rotating shallow-water equations on
  the sphere.
\newblock \emph{Quarterly Journal of the Royal Meteorological Society},
  145\penalty0 (720):\penalty0 1070--1088, 2019.

\bibitem[Brecht et~al.(2021)Brecht, Li, Bauer, and
  M{\'e}min]{rudigerbrecht_2021_4884919}
R{\"u}diger Brecht, Long Li, Werner Bauer, and Etienne M{\'e}min.
\newblock Rudigerbrecht/rsw-lu: First release, May 2021.
\newblock URL \url{https://doi.org/10.5281/zenodo.4884919}.

\bibitem[Buizza et~al.(1999)Buizza, Miller, and Palmer]{Buizza99}
R.~Buizza, M.~Miller, and T.N. Palmer.
\newblock Stochastic representation of model uncertainties in the {ECMWF}
  ensemble prediction system.
\newblock \emph{Quarterly Journal Royal Meteorological Society}, 125:\penalty0
  2887--2908, 1999.

\bibitem[Chandramouli et~al.(2020)Chandramouli, Memin, and
  Heitz]{Chandramouli-JCP-20}
P.~Chandramouli, E.~Memin, and D.~Heitz.
\newblock 4{D} large scale variational data assimilation of a turbulent flow
  with a dynamics error model.
\newblock \emph{Journal of Computational Physics}, 412:\penalty0 109446, 2020.

\bibitem[Chapron et~al.(2018)Chapron, D\'{e}rian, M\'{e}min, and
  Resseguier]{Chapron2018}
B.~Chapron, P.~D\'{e}rian, E.~M\'{e}min, and V.~Resseguier.
\newblock Large-scale flows under location uncertainty: a consistent stochastic
  framework.
\newblock \emph{Quarterly Journal of the Royal Meteorological Society},
  144\penalty0 (710):\penalty0 251--260, 2018.

\bibitem[Chen et~al.(2011)Chen, Gunzburger, and Ringler]{chen11Ay}
Q.~Chen, M.~Gunzburger, and T.~Ringler.
\newblock A scale-invariant formulation of the anticipated potential vorticity
  method.
\newblock \emph{Monthly Weather Review}, 139\penalty0 (8):\penalty0 2614--2629,
  2011.

\bibitem[Da~Prato and Zabczyk(2014)]{DaPrato2014}
G.~Da~Prato and J.~Zabczyk.
\newblock \emph{Stochastic equations in infinite dimensions}.
\newblock Encyclopedia of Mathematics and its Applications. Cambridge
  University Press, 2 edition, 2014.

\bibitem[Desbrun et~al.(2014)Desbrun, Gawlik, Gay-Balmaz, and
  Zeitlin]{desbrun2014variational}
M.~Desbrun, E.S. Gawlik, F.~Gay-Balmaz, and V.~Zeitlin.
\newblock Variational discretization for rotating stratified fluids.
\newblock \emph{Discrete \& Continuous Dynamical Systems-A}, 34\penalty0
  (2):\penalty0 477, 2014.

\bibitem[Frank et~al.(2003)Frank, Gottwald, and Reich]{Frank2003}
J.~Frank, G.~Gottwald, and S.~Reich.
\newblock \emph{A {H}amiltonian Particle-Mesh Method for the Rotating
  Shallow-Water Equations}, volume~26 of \emph{Lecture Notes in Computational
  Science and Engineering}, pages 131--142.
\newblock Springer, meshless methods for partial differential equations
  edition, 07 2003.

\bibitem[Franzke et~al.(2006)Franzke, Majda, and Vanden-Eijnden]{Franzke2005}
C.~Franzke, A.~Majda, and E.~Vanden-Eijnden.
\newblock Low-order stochastic mode reduction for a realistic barotropic model
  climate.
\newblock \emph{Journal of the Atmospheric Sciences}, 62\penalty0 (6):\penalty0
  1722--1757, 2006.

\bibitem[Franzke and Majda(2006)]{Franzke2006}
C.~E. Franzke and A.~J. Majda.
\newblock Low-order stochastic mode reduction for a prototype atmospheric
  {GCM}.
\newblock \emph{Journal of the Atmospheric Sciences}, 63\penalty0 (2):\penalty0
  457--479, 2006.

\bibitem[Franzke et~al.(2015)Franzke, O'Kane, Berner, Williams, and
  Lucarini]{Franzke2015}
C.~E. Franzke, T.~J. O'Kane, J.~Berner, P.~D. Williams, and V.~Lucarini.
\newblock Stochastic climate theory and modeling.
\newblock \emph{Wiley Interdisciplinary Reviews: Climate Change}, 6\penalty0
  (1):\penalty0 63--78, 2015.

\bibitem[Frederiksen et~al.(2013)Frederiksen, O'Kane, and
  Zidikheri]{frederiksen2013}
J.~S. Frederiksen, T.~J. O'Kane, and M.~J. Zidikheri.
\newblock Subgrid modelling for geophysical flows.
\newblock \emph{Philosophical Transactions of the Royal Society A:
  Mathematical, Physical and Engineering Sciences}, 371\penalty0
  (1982):\penalty0 20120166, 2013.

\bibitem[Galewsky et~al.(2004)Galewsky, Scott, and Polvani]{Galewsky2004}
J.~Galewsky, R.~K. Scott, and L.~M. Polvani.
\newblock An initial-value problem for testing numerical models of the global
  shallow-water equations.
\newblock \emph{Tellus A: Dynamic Meteorology and Oceanography}, 56\penalty0
  (5):\penalty0 429--440, 2004.

\bibitem[Gawlik et~al.(2011)Gawlik, Mullen, Pavlov, Marsden, and
  Desbrun]{gawlik2011geometric}
E.S. Gawlik, P.~Mullen, D.~Pavlov, J.E. Marsden, and M.~Desbrun.
\newblock Geometric, variational discretization of continuum theories.
\newblock \emph{Physica D: Nonlinear Phenomena}, 240\penalty0 (21):\penalty0
  1724--1760, 2011.

\bibitem[Gent and McWilliams(1990)]{Gent1990}
P.~R. Gent and J.~C. McWilliams.
\newblock Isopycnal mixing in ocean circulation models.
\newblock \emph{Journal of Physical Oceanography}, 20\penalty0 (1):\penalty0
  150--155, 1990.

\bibitem[Gent et~al.(1995)Gent, Willebrand, McDougall, and
  McWilliams]{Gent1995}
P.~R. Gent, J.~Willebrand, T.~J. McDougall, and J.~C. McWilliams.
\newblock Parameterising eddy-induced tracer transports in ocean circulation
  models.
\newblock \emph{Journal of Physical Oceanography}, 25:\penalty0 463--474, 1995.

\bibitem[Gottwald and Harlim(2013)]{Gottwald2013}
G.~Gottwald and J.~Harlim.
\newblock The role of additive and multiplicative noise in filtering complex
  dynamical systems.
\newblock \emph{Proceedings of the Royal Society A: Mathematical, Physical and
  Engineering Science}, 469\penalty0 (2155):\penalty0 20130096, 2013.

\bibitem[Gottwald et~al.(2017)Gottwald, Crommelin, and Franzke]{Gottwald2017}
G.~Gottwald, D.~T. Crommelin, and C.~E. Franzke.
\newblock Stochastic climate theory.
\newblock In \emph{Nonlinear and Stochastic Climate Dynamics}, pages 209--240.
  Cambridge University Press, 2017.

\bibitem[Griffies(1998)]{Griffies1998}
S.~M. Griffies.
\newblock The {G}ent-{M}c{W}illiams skew flux.
\newblock \emph{Journal of Physical Oceanography}, 28\penalty0 (5):\penalty0
  831--841, 1998.

\bibitem[Gugole and Franzke(2019)]{Gugole2019}
F.~Gugole and C.~E. Franzke.
\newblock Numerical development and evaluation of an energy conserving
  conceptual stochastic climate model.
\newblock \emph{Mathematics of Climate and Weather Forecasting}, 5\penalty0
  (1):\penalty0 45--64, 2019.

\bibitem[Hairer et~al.(2006)Hairer, Lubich, and Wanner]{hairer2006geometric}
E.~Hairer, C.~Lubich, and G.~Wanner.
\newblock \emph{Geometric numerical integration: structure-preserving
  algorithms for ordinary differential equations}, volume~31.
\newblock Springer Science \& Business Media, 2006.

\bibitem[Hamill(2001)]{Hamill2001}
T.~M. Hamill.
\newblock Interpretation of rank histograms for verifying ensemble forecasts.
\newblock \emph{Monthly Weather Review}, 129:\penalty0 550--560, 2001.

\bibitem[Hasselmann(1976)]{Hasselmann-76}
K.~Hasselmann.
\newblock Stochastic climate models part {I}. theory.
\newblock \emph{Tellus}, 28:\penalty0 473--485, 1976.

\bibitem[Hecht et~al.(2008)Hecht, Holm, Petersen, and Wingate]{Hecht08}
M.~Hecht, D.~Holm, M.~Petersen, and B.~Wingate.
\newblock Implementation of the {L}ans-alpha turbulence model in a primitive
  equation ocean model.
\newblock \emph{Journal of Computational Physics}, 27\penalty0 (11):\penalty0
  5691--5711, 2008.

\bibitem[Holm(2015)]{Holm-15}
D.D. Holm.
\newblock Variational principles for stochastic fluid dynamics.
\newblock \emph{Proceedings of the Royal Society A: Mathematical, Physical and
  Engineering Science}, 471\penalty0 (20140963), 2015.

\bibitem[Hsu(2002)]{Hsu2002}
E.P. Hsu.
\newblock \emph{Stochastic Analysis on Manifolds}.
\newblock Graduate studies in mathematics. American Mathematical Society, 2002.

\bibitem[Kadri~Harouna and M\'{e}min(2017)]{Kadri2017}
S.~Kadri~Harouna and E.~M\'{e}min.
\newblock Stochastic representation of the {R}eynolds transport theorem:
  revisiting large-scale modeling.
\newblock \emph{Computers and Fluids}, 156:\penalty0 456--469, 2017.

\bibitem[Kafiabad et~al.(2021)Kafiabad, Vanneste, and
  Young]{kafiabad_vanneste_young_2021}
H.~A. Kafiabad, J.~Vanneste, and W.~R. Young.
\newblock Wave-averaged balance: a simple example.
\newblock \emph{Journal of Fluid Mechanics}, 911:\penalty0 R1, 2021.
\newblock \doi{10.1017/jfm.2020.1032}.

\bibitem[Kjellsson and Zanna(2017)]{Kjellsson2017}
J.~Kjellsson and L.~Zanna.
\newblock The impact of horizontal resolution on energy transfers in global
  ocean models.
\newblock \emph{Fluids}, 2\penalty0 (3):\penalty0 45, 2017.

\bibitem[Kloeden and Platen(1992)]{Kloeden1992}
P.~E. Kloeden and E.~Platen.
\newblock \emph{Numerical Solution of Stochastic Differential Equations},
  volume~23.
\newblock Springer-Verlag Berlin Heidelberg, 1992.

\bibitem[Kunita(1997)]{Kunita1997}
H.~Kunita.
\newblock \emph{Stochastic flows and stochastic differential equations},
  volume~24 of \emph{Cambridge Studies in Advanced Mathematics}.
\newblock Cambridge University Press, 1997.

\bibitem[Leimkuhler and Reich(2004)]{leim04Ay}
B.~Leimkuhler and S.~Reich.
\newblock \emph{{Simulating Hamiltonian dynamics}}.
\newblock Cambridge University Press, Cambridge, 2004.

\bibitem[Leith(1975)]{Leith75}
C.~Leith.
\newblock Climate response and fluctuation dissipation.
\newblock \emph{Journal of the Atmospheric Sciences}, 32\penalty0
  (10):\penalty0 2022--2026, 1975.

\bibitem[Leith(1990)]{Leith90}
C.~Leith.
\newblock Stochastic backscatter in a subgrid-scale model: plane shear mixing
  layer.
\newblock \emph{Physics of Fluids}, 2\penalty0 (3):\penalty0 1521--1530, 1990.

\bibitem[L\'{e}vy et~al.(2010)L\'{e}vy, Klein, Tr\'{e}guier, Iovino, Madec,
  Masson, and Takahashi]{Levy2010}
M.~L\'{e}vy, P.~Klein, A.~M. Tr\'{e}guier, D.~Iovino, G.~Madec, S.~Masson, and
  K.~Takahashi.
\newblock Modifications of gyre circulation by sub-mesoscale physics.
\newblock \emph{Ocean Modelling}, 34\penalty0 (1-2):\penalty0 1--15, 2010.

\bibitem[L\'{e}vy et~al.(2012)L\'{e}vy, Resplandy, Klein, Capet, Iovino, and
  Eth'{e}]{Levy2012}
M.~L\'{e}vy, L.~Resplandy, P.~Klein, X.~Capet, D.~Iovino, and C.~Eth'{e}.
\newblock Grid degradation of submesoscale resolving ocean models: Benefits for
  offline passive tracer transport.
\newblock \emph{Ocean Modelling}, 48\penalty0 (1-2):\penalty0 1--9, 2012.

\bibitem[Lorenz(1963)]{Lorenz63}
E.~Lorenz.
\newblock Deterministic nonperiodic flow.
\newblock \emph{Journal of the Atmospheric Sciences}, 73\penalty0
  (12):\penalty0 130--141, 1963.

\bibitem[Majda et~al.(1999)Majda, Timofeyev, and Eijnden]{Majda99}
A.~Majda, I.~Timofeyev, and E.~Vanden Eijnden.
\newblock Models for stochastic climate prediction.
\newblock \emph{Proceedings of the National Academy of Sciences of the United
  States of America}, 1999.

\bibitem[Majda et~al.(2008)Majda, Franzke, and Khouider]{Majda2008}
A.~Majda, C.~Franzke, and B.~Khouider.
\newblock An applied mathematics perspective on stochastic modelling for
  climate.
\newblock \emph{Philosophical Transactions of the Royal Society of London A:
  Mathematical, Physical and Engineering Sciences}, 366\penalty0
  (1875):\penalty0 2427--2453, 2008.

\bibitem[Marsden and West(2001)]{marsden2001discrete}
J.~E. Marsden and M.~West.
\newblock Discrete mechanics and variational integrators.
\newblock \emph{Acta Numerica}, 10\penalty0 (1):\penalty0 357--514, 2001.

\bibitem[Mason and Thomson(1992)]{Mason92}
P.J. Mason and D.J. Thomson.
\newblock Stochastic backscatter in large-eddy simulations of boundary layers.
\newblock \emph{Journal of Fluid Mechanics}, 242:\penalty0 51--78, 1992.

\bibitem[McRae and Cotter(2014)]{mcrae2014energy}
A.~TT. McRae and C.J. Cotter.
\newblock Energy-and enstrophy-conserving schemes for the shallow-water
  equations, based on mimetic finite elements.
\newblock \emph{Quarterly Journal of the Royal Meteorological Society},
  140\penalty0 (684):\penalty0 2223--2234, 2014.

\bibitem[McWilliams et~al.(2004)McWilliams, Restrepo, and
  Lane]{Mc-Williams-Restrepo-Lane-04}
J.~McWilliams, J.~Restrepo, and E.~Lane.
\newblock An asymptotic theory for the interaction of waves and currents in
  coastal waters.
\newblock \emph{Journal of Fluid Mechanics}, 511:\penalty0 135--178, 2004.

\bibitem[McWilliams(2006)]{McWilliams2006GFD}
J.~C. McWilliams.
\newblock \emph{Fundamentals of Geophysical Fluid Dynamics}.
\newblock Cambridge University Press, 2006.

\bibitem[Mellor(2016)]{Mellor2016}
George Mellor.
\newblock On theories dealing with the interaction of surface waves and ocean
  circulation.
\newblock \emph{Journal of Geophysical Research: Oceans}, 121\penalty0
  (7):\penalty0 4474--4486, 2016.
\newblock \doi{https://doi.org/10.1002/2016JC011768}.
\newblock URL
  \url{https://agupubs.onlinelibrary.wiley.com/doi/abs/10.1002/2016JC011768}.

\bibitem[M\'{e}min(2014)]{Memin2014}
E.~M\'{e}min.
\newblock Fluid flow dynamics under location uncertainty.
\newblock \emph{Geophysical \& Astrophysical Fluid Dynamics}, 108\penalty0
  (2):\penalty0 119--146, 2014.

\bibitem[Meneveau and Katz(2000)]{Meneveau00}
C.~Meneveau and J.~Katz.
\newblock Scale-invariance and turbulence models for large-eddy simulation.
\newblock \emph{Annual Review of Fluid Mechanics}, 32:\penalty0 1--32, 2000.

\bibitem[Palmer and Williams(2008)]{Palmer08}
T.~Palmer and P.~Williams.
\newblock Theme issue 'stochastic physics and climate modelling'.
\newblock \emph{Philosophical Transactions of the Royal Society A:
  Mathematical, Physical and Engineering Sciences}, 366\penalty0 (1875), 2008.

\bibitem[Pavlov et~al.(2011)Pavlov, Mullen, Tong, Kanso, Marsden, and
  Desbrun]{pavl11a}
D.~Pavlov, P.~Mullen, Y.~Tong, E.~Kanso, J.~E. Marsden, and M.~Desbrun.
\newblock Structure-preserving discretization of incompressible fluids.
\newblock \emph{Physica D: Nonlinear Phenomena}, 240\penalty0 (6):\penalty0
  443--458, 2011.

\bibitem[Perot et~al.(2006)Perot, Vidovic, and Wesseling]{perot2006mimetic}
J.~B. Perot, D.~Vidovic, and P.~Wesseling.
\newblock Mimetic reconstruction of vectors.
\newblock In \emph{Compatible Spatial Discretizations}, pages 173--188.
  Springer, 2006.

\bibitem[Pope(2000)]{Pope2000}
S.~Pope.
\newblock \emph{Turbulent flows}.
\newblock Cambridge University Press, 2000.

\bibitem[Porta~Mana and Zanna(2014)]{PortaMana-14}
P.~Porta~Mana and L.~Zanna.
\newblock Toward a stochastic parametrization of ocean mesoscale eddies.
\newblock \emph{Ocean Modelling}, 79\penalty0 (1-20), 2014.

\bibitem[Resseguier et~al.(2017{\natexlab{a}})Resseguier, M\'{e}min, and
  Chapron]{Resseguier2017gafd1}
V.~Resseguier, E.~M\'{e}min, and B.~Chapron.
\newblock Geophysical flows under location uncertainty, part {I}: {R}andom
  transport and general models.
\newblock \emph{Geophysical \& Astrophysical Fluid Dynamics}, 111\penalty0
  (3):\penalty0 149--176, 2017{\natexlab{a}}.

\bibitem[Resseguier et~al.(2017{\natexlab{b}})Resseguier, M\'{e}min, and
  Chapron]{Resseguier2017gafd2}
V.~Resseguier, E.~M\'{e}min, and B.~Chapron.
\newblock Geophysical flows under location uncertainty, part {II}:
  {Q}uasi-geostrophic models and efficient ensemble spreading.
\newblock \emph{Geophysical \& Astrophysical Fluid Dynamics}, 111\penalty0
  (3):\penalty0 177--208, 2017{\natexlab{b}}.

\bibitem[Resseguier et~al.(2017{\natexlab{c}})Resseguier, M\'{e}min, and
  Chapron]{Resseguier2017gafd3}
V.~Resseguier, E.~M\'{e}min, and B.~Chapron.
\newblock Geophysical flows under location uncertainty, part {III}: {SQG} and
  frontal dynamics under strong turbulence.
\newblock \emph{Geophysical \& Astrophysical Fluid Dynamics}, 111\penalty0
  (3):\penalty0 209--227, 2017{\natexlab{c}}.

\bibitem[Resseguier et~al.(2020)Resseguier, Li, Jouan, Derian, M\'{e}min, and
  Chapron]{Resseguier2020arcme}
V.~Resseguier, L.~Li, G.~Jouan, P.~Derian, E.~M\'{e}min, and B.~Chapron.
\newblock New trends in ensemble forecast strategy: uncertainty quantification
  for coarse-grid computational fluid dynamics.
\newblock \emph{Archives of Computational Methods in Engineering}, pages
  1886--1784, 2020.

\bibitem[Ringler and Randall(2002)]{ringler2002potential}
T.D. Ringler and D.A. Randall.
\newblock A potential enstrophy and energy conserving numerical scheme for
  solution of the shallow-water equations on a geodesic grid.
\newblock \emph{Monthly Weather Review}, 130\penalty0 (5):\penalty0 1397--1410,
  2002.

\bibitem[Salmon(2013)]{Salmon2013}
R.~Salmon.
\newblock An alternative view of generalized {L}agrangian mean theory.
\newblock \emph{Journal of Fluid Mechanics}, 719\penalty0 (165-182), 2013.

\bibitem[Schmid(2010)]{Schmid10}
P.~Schmid.
\newblock Dynamic mode decomposition of numerical and experimental data.
\newblock \emph{Journal of Fluid Mechanics}, 656:\penalty0 5--28, 2010.

\bibitem[Shutts(2005)]{Shutts05}
G.~Shutts.
\newblock A kinetic energy backscatter algorithm for use in ensemble prediction
  systems.
\newblock \emph{Quarterly Journal of the Royal Meteorological Society},
  612:\penalty0 3079--3012, 2005.

\bibitem[Slingo and Palmer(2011)]{Slingo11}
J.~Slingo and T.~Palmer.
\newblock Uncertainty in weather and climate prediction.
\newblock \emph{Philosophical Transactions of the Royal Society A:
  Mathematical, Physical and Engineering Sciences}, 369:\penalty0 4751--4767,
  2011.

\bibitem[Talagrand et~al.(1997)Talagrand, Vautard, and Strauss]{Talagrand1997}
O.~Talagrand, R.~Vautard, and B.~Strauss.
\newblock Evaluation of probabilistic prediction systems.
\newblock Workshop on Predictability, ECMWF, 1997.

\bibitem[Vallis(2017)]{Vallis2017}
G.~K. Vallis.
\newblock \emph{Atmospheric and oceanic fluid dynamics: fundamentals and
  large-scale circulation}.
\newblock Cambridge University Press, 2 edition, 2017.

\bibitem[Weigel(2012)]{Weigel2012}
A.~P. Weigel.
\newblock Ensemble forecasts.
\newblock In \emph{Forecast Verification}, chapter~8, pages 141--166. John
  Wiley and Sons, Ltd, 2012.

\bibitem[Xie and Vanneste(2015)]{Xie2015}
J.-H. Xie and J.~Vanneste.
\newblock A generalised-{L}agrangian-mean model of the interactions between
  near-inertial waves and mean flow.
\newblock \emph{Journal of Fluid Mechanics}, 774:\penalty0 143--169, 2015.

\bibitem[Young and Jelloul(1997)]{Young97}
W.~Young and M.~Ben Jelloul.
\newblock Propagation of near-inertial oscillation through a geostrophic flow.
\newblock \emph{Journal of Marine Research}, 55\penalty0 (4):\penalty0
  735--766, 1997.

\end{thebibliography}

\end{document}